\documentclass{revtex4}
\usepackage{amsthm,amssymb,amsmath}				
\usepackage{graphicx,comment}
\usepackage{float}   
 \usepackage{xcolor}
 \usepackage{color}
 \definecolor{mycolor}{rgb}{0.8, 0.2, 0.2}

\begin{document}

\title{KG- oscillators in a spinning cosmic string spacetime and an external magnetic field}
\author{Omar Mustafa}
\email{omar.mustafa@emu.edu.tr}
\affiliation{Department of Physics, Eastern Mediterranean University, 99628, G. Magusa,
north Cyprus, Mersin 10 - Turkiye.}
\author{Abdullah Guvendi}
\email{abdullah.guvendi@erzurum.edu.tr}
\affiliation{Department of Basic Sciences, Erzurum Technical University, 25050, Erzurum, T\"{u}rkiye.}

\begin{abstract}
\textbf{Abstract:} We study the Klein-Gordon (KG) oscillators in a spinning cosmic string spacetime and an external magnetic field. The corresponding KG-equation is shown to admit a solution in the form of the confluent hypergeometric functions/polynomials. Consequently, the corresponding energies are shown to be given in a quadratic equation of delicate nature that has to be solved in an orderly manner (for it involves the energies for KG-particles/antiparticles, $E=E_{\pm}=\pm\left\vert E\right\vert $ along with the magnetic quantum number $m=m_{\pm }=\pm |m|$). Following a case-by-case strategy allowed us to clearly observe the effects of the spinning cosmic string on the spectroscopic structure of the KG-oscillators.  Interestingly, we have observed that the coexistence of a spinning cosmic string and an external magnetic field eliminates the effect of the wedge parameter for KG-particles  ($E=E_{+}$), with $m=m_{+}$, but not for the KG-antiparticles  ($E=E_{-}$), with $m=m_{-}$. Such coexistence is observed to break the symmetry of the energies of the KG-particles and the antiparticles about $E=0$  for the KG-oscillators. However, for KG-particles ($E=E_{+}$), with $m=m_{-}$, and KG-antiparticles ($E=E_{-}$), with $m=m_{+}$, they are found to be unfortunate because they are indeterminable. Moreover, for the spinning parameter $\beta >>1$, the clustering of the energy levels is observed eminent to indicate that there is no distinction between energy levels at such values of $\beta $.

\textbf{PACS }numbers\textbf{: }05.45.-a, 03.50.Kk, 03.65.-w

\textbf{Keywords:} Klein-Gordon oscillators, spinning cosmic string spacetime, external magnetic field, Landau levels.
\end{abstract}

\maketitle

\section{Introduction}

Kibble \cite{1.1} has suggested that cosmic strings find existential support in modern superstring theories, both in compactification models and in extended additional-dimensional models. The eventual discovery of which would be of a significant contribution for cosmological and fundamental physics. Cosmic strings, among other topological defects in the spacetime fabric, are one-dimensional stable configurations of matter that are formed at cosmological phase transition in the very early universe. The study of which provide crucial links in the understanding of the physics of the evolving universe \cite{1.1,1.2,1.3,1.4,1.5}. Their gravitational fields' feasible role in the galaxy formation makes them of particular interest \cite%
{1.4,1.5}. Cosmic strings (spinning or static) are shown to introduce interesting effects like self-force particles \cite{1.6,1.7}, gravitational lensing \cite{1.8}, and high-energy particles \cite{1.9,1.10,1.11}, Their most effective detection is contemplated to be through their gravitational lensing fingerprints/signatures \cite{1.8} that are different from those introduced by standard lenses (i.e. compact clumps of matter). However, the discovery of the peculiar object CSL-1 has stimulated strong arguments among the physics community, in favor of being the first case of gravitational lensing by a cosmic string \cite{1.8}.

Cosmic strings (spinning or static) share some similar characteristics represented (in $\hbar =c=1$ units, to be used throughout) by the so called wedge parameter $\alpha =1-4\eta G$, which is a measure of angle deficit
produced by the string, where $G$ is the gravitational Newton constant and $%
\eta $ is the linear mass density of the string. A spinning cosmic string has an additional characteristic represented by a rotational/spinning
parameter $\beta =4GJ$, where $\beta $ has the unit of length and $J$ \ is the angular momentum per unit length. The line element metric that describes the exterior and interior spacetime structure generated by a spinning
cosmic string reads%
\begin{equation}
ds^{2}=-\left( A\,dt+B\,d\varphi \right) ^{2}+d\rho ^{2}+C^{2}\,d\varphi
^{2}+dz^{2},  \label{1.1}
\end{equation}%
where $A$, $B$, and $C$ are functions of the radial coordinate $\rho$ only \cite%
{1.12}. The Lorentz invariance along the $z$-axis notably suggests that the solution of such a (3+1)-dimensional field equation may very well be
interpreted as a solution for the corresponding (2+1)-dimensional one \cite%
{1.12,1.13}. Moreover, for the case $A=1$, $B=\beta $, and $C=\alpha \left(
\rho +\rho _{\circ }\right) =\alpha \,r$, this metric is a flat vacuum solution \cite{1.14}, without the interior structure, and reads%
\begin{equation}
ds^{2}=-dt^{2}-2\beta \,dt\,d\varphi +\left( \alpha ^{2}r^{2}-\beta
^{2}\right) \,d\varphi ^{2}+dr^{2}\,+dz^{2}.  \label{1.2}
\end{equation}%
\textcolor{blue}{Here, $\rho _{\circ }$ is a constant that describes the origin of the exterior radial coordinate where the radial coordinates of the interior and exterior structures coincide \cite{1.12}. Further other solutions representing spinning cosmic strings are also reported in \cite{1.15,1.16,1.17,1.18}.  At this point, one should notice that $\beta\neq 0$ induces a helical structure of time and $\alpha < 1$ produces a conical topology. One should, moreover, observe that $g_{\varphi\varphi}$ being positive  would suggest a restriction on the radial coordinate $r\geq\beta/\alpha\Rightarrow r_{min}=\beta/\alpha$, which, in fact, describes the origin of the exterior radial coordinate, i.e., the region of interest of the current study.  One should also note that $r=\rho+\rho_{\circ}\geq\rho_{\circ}$ would immediately suggest that $r_{min}=\rho_{\circ}=\beta/\alpha$. Consequently, there are no regions to be classified as forbidden regions for such a spacetime (\ref{1.2}) structure. An argument like "\textit{if $r< \beta/\alpha$ results in $g_{\varphi\varphi}$ becoming negative to classify a forbidden region of spacetime}"  \cite{1.181} cannot be valid because $r=\rho+\rho_{\circ}\Rightarrow \rho+\beta/\alpha\nless \beta/\alpha$ since $[0,\rho_{\circ}]\ni\rho\nless 0$, (the range of the coordinate of the interior structure).}

In the current methodical proposal, we shall consider KG-oscillators in a spinning cosmic string described by (\ref{1.2}), where, the corresponding metric tensors are given by%
\begin{equation}
g_{\mu \nu }=\left( 
\begin{tabular}{cccc}
$-1$ & $\,\,0$ & $-\beta $ & $0$ \\ 
$0$ & $\,\,1$ & $0$ & $0$ \\ 
$-\beta $ & $\,\,0$ & $\alpha ^{2}r^{2}-\beta ^{2}$ & $0$ \\ 
$0$ & $\,\,0$ & $0$ & $1$%
\end{tabular}%
\right) ;\;\det \left( g_{\mu \nu }\right) =g=-\alpha ^{2}r^{2},\;g^{\mu \nu
}=\left( 
\begin{tabular}{cccc}
$\frac{\beta ^{2}-\alpha ^{2}r^{2}}{\alpha ^{2}r^{2}}$ & $0$ & $-\frac{\beta 
}{\alpha ^{2}r^{2}}$ & $0$ \\ 
$0$ & $1$ & $0$ & $0$ \\ 
$-\frac{\beta }{\alpha ^{2}r^{2}}$ & $0$ & $\frac{1}{\alpha ^{2}r^{2}}$ & $0$
\\ 
$0$ & $0$ & $0$ & $1$%
\end{tabular}%
\right)   \label{1.3}
\end{equation}
In the literature, however, we have observed that only two distinct studies were carried out on the Landau levels of relativistic and/or non-relativistic quantum particles in a spinning cosmic string spacetime 
\cite{1.19,1.20,1.201}, unlike the so many such studies of quantum particles in static cosmic string spacetime \cite%
{1.21,1.22,1.23,1.24,1.25,1.26,1.27,1.28,1.29,1.30}. Motivated by such a few studies and the fundamental pedagogical interest on the KG-oscillators for quantum gravity, we intend to fill, at least partially, this gap and report some interesting gravitational field effects of a spinning cosmic string on the spectroscopic structure of KG-oscillators in an external magnetic field. As a special case of the current study, however, we shall also show that the results reported by Cunha et al. \cite{1.19} (namely, their Eq.s (15) and (16)) describe only a set of Landau levels for KG-particles (i.e., $E=E_{+}=|E|$) with positive magnetic quantum numbers (i.e. $\ell =\ell _{+}=+|\ell |$ as used by Cunha et al. \cite{1.19}, here we shall use the textbook magnetic quantum number $m=m_{+}=+|m|$ instead), and KG-antiparticles (i.e., 
$E=E_{-}=-|E|$) with negative magnetic quantum numbers (i.e., $\ell =\ell
_{-}=-|\ell |$ which would correspond to $m=m_{-}=-|m|$ in the current study). Moreover, the very fundamental textbook nature of 
Schr\"{o}dinger, Dirac, and KG oscillators problems (as being corner stones for quantum mechanics) makes the study of the effects of gravitational fields, introduced by different spacetimes backgrounds, on their spectroscopic structure crucial and interesting for quantum gravity as well \cite%
{1.31,1.32,1.33,1.34,1.35,1.36,1.37,1.38,1.39,1.40,1.41,1.42,1.43,1.44,1.45,1.46}%
, to mention a few. The above mentioned inspires and motivates the current study.

Our study is organized as follows. In Section 2, we discuss the KG-oscillators in a spinning cosmic string spacetime and an external magnetic field. We start with the KG-equation in such a spacetime background and
bring it to the one-dimensional Schr\"{o}dinger-oscillator form (hence the notion KG-oscillators is rendered unavoidable) that admits a solution in the form of confluent hypergeometric functions/polynomials. Consequently, the corresponding energies are given in terms of a quadratic equation of a delicate nature, as it includes $E=E_{\pm }=\pm |E|$ (i.e., energies for KG-particles, $E=E_{+}=\left\vert E\right\vert $, and antiparticles, $%
E=E_{-}=-\left\vert E\right\vert $) and the magnetic quantum number $%
m=m_{\pm }=\pm |m|=0,\pm 1,\pm 2,\cdots $. This in turn adds more complexity to the problem at hand (see the result in (\ref{2.8}) below). Moreover, we retrieve the exact results for the special cases (of the more general one used here) considered by Medeiros et al. \cite{1.24} and by Cunha et al. \cite{1.19}. We elaborate on the validity of the exact solutions reported by Cunha et al. \cite{1.19}. We go around such complexity solving for case-by-case and discuss KG-oscillators in a spinning cosmic string spacetime in an external magnetic field for $S$-states (i.e., $m=0$) in section 2.A. In section 2.B, we report for KG-particles, $%
E=E_{+}=\left\vert E\right\vert $, associated with $m=m_{+}$ and for KG-antiparticles associated with $m=m_{\_}$. In Section 2.C, we report that the KG-oscillators with energies $E=E_{\pm }$ associated with $m=m_{\mp }$, respectively, are left unfortunate (in the sense that their energies cannot be determined). Interesting effects of the spinning cosmic string are observed. Namely, for the spinning parameter $\beta >>1$ clustering of the energy levels is observed eminent, which would, in turn, indicate that there is no distinction between energy levels at such values of $\beta $.  Moreover, we observe that whilst the Landau-like energies of the KG-particles ($E=E_{+}$), with $m=m_{+}$, have no explicit dependence on the spinning string parameter $\beta$ or the wedge parameter $\alpha$, the KG-antiparticles ($E=E_{-}$), with $m=m_{-}$, have such an explicit dependence. Interestingly, therefore, the coexistence of a spinning cosmic string and an external magnetic field is observed to eliminate/cancel the effect of the wedge parameter $\alpha$ ( which is readily a byproduct of the cosmic string) for KG-particles  ($E=E_{+}$), with $m=m_{+}$, but not for the KG-antiparticles  ($E=E_{-}$), with $m=m_{-}$.We conclude in Section 3.

\section{KG-oscillators in a spinning cosmic string spacetime in an external
magnetic field}

The KG-oscillators are described by the KG-equation 
\begin{equation}
\left( \frac{1}{\sqrt{-g}}\tilde{D}^+_{\mu }\sqrt{-g}g^{\mu \nu }\tilde{D}^-%
_{\nu }\right) \,\Psi \left( t,r,\varphi ,z\right) =m_{\circ }^{2}\,\Psi
\left( t,r,\varphi ,z\right) ,  \label{2.1}
\end{equation}%
where $\tilde{D}^\pm_{\mu }=D_{\mu }\pm\mathcal{F}_{\mu }$ is in a non-minimal coupling form with $\mathcal{F}_{\mu }$ $\in 
\mathbb{R}
$, $D_{\mu }=\partial _{\mu }-ieA_{\mu }$ is the gauge-covariant derivative that admits minimal coupling, $A_{\nu }=\left( 0,0,A_{\varphi },0\right) $ is the 4-vector potential, and $m_{\circ }$ is the rest mass energy (i.e., $%
m_{\circ }\equiv m_{\circ }c^{2}$, with $\hbar =c=1$ units to be used throughout). Incorporating the KG oscillators \cite{1.31,1.32} in the process would be done using $\mathcal{F}_{\mu }=\left( 0,\mathcal{F}_{r},0,0\right)
;\,\mathcal{F}_{r}=\Omega r$. Under such settings, equation (\ref{2.1}) would read%
\begin{equation}
\left\{ \frac{1}{r}\left( \partial _{r}+\mathcal{F}_{r}\right) \,r\,\left(
\partial _{r}-\mathcal{F}_{r}\right) -\frac{\tilde{m}^{2}}{r^{2}}-\tilde{B}%
^{2}r^{2}+\mathcal{E}^{2}\right\} R\left( r\right) =0,  \label{2.2}
\end{equation}%
where we have used 
\begin{equation}
\Psi \left( t,r,\varphi ,z\right) =R\left( r\right) e^{i\left( m\varphi
+kz-Et\right) },\;A_{\varphi }=\frac{\alpha B}{2}r^{2},\;\tilde{B}=\frac{eB}{%
2}  \label{2.3}
\end{equation}%
and%
\begin{equation}
\tilde{m}^{2}=\left( \frac{\beta E+m}{\alpha }\right) ^{2},\;\mathcal{E}%
^{2}=E^{2}+\frac{meB}{\alpha }+\frac{eB}{\alpha }\beta E-k^{2}-m_{\circ
}^{2}.  \label{2.4}
\end{equation}%
Equation (\ref{2.2}), with $\mathcal{F}_{r}=\Omega r$ and $R\left( r\right)
=U(r)/\sqrt{r}$\ would yield%
\begin{equation}
\left\{ \partial _{r}^{2}-\frac{\left( \tilde{m}^{2}-1/4\right) }{r^{2}}-%
\tilde{\Omega}^{2}r^{2}+\mathcal{\tilde{E}}^{2}\right\} U\left( r\right)
=0,\;\tilde{\Omega}^{2}=\tilde{B}^{2}+\Omega ^{2},\;\mathcal{\tilde{E}}^{2}=%
\mathcal{E}^{2}-2\Omega .  \label{2.5}
\end{equation}%
\textcolor{blue}{It is obvious that the effective potential $$V_{eff}(r)=\frac{\left( \tilde{m}^{2}-1/4\right) }{r^{2}}+%
\tilde{\Omega}^{2}r^{2}$$ manifestly introduced by the spinning cosmic string spacetime does not generate an infinite repulsive hard wall (i.e., domain wall) at $r=\rho_{\circ}=\beta/\alpha$. Therefore, one should not expect the corresponding radial wave function $U(r)$ to vanish at $r=\beta/\alpha$.  Under such settings, our radial equation (\ref{2.5}) would admit a solution in the form of confluent hypergeometric functions, so that%
\begin{equation}
R(r)=\frac{U\left( r\right)}{\sqrt{r}} =\mathcal{C}\,r^{|\tilde{m}|}e^{-\tilde{\Omega}%
r^{2}/2}\,_{1}F_{1}\left( \frac{1}{2}+\frac{|\tilde{m}|}{2}-\frac{\mathcal{%
\tilde{E}}^{2}}{4\tilde{\Omega}},|\tilde{m}|+1,\tilde{\Omega}r^{2}\right),
\label{2.6}
\end{equation}%
where $\mathcal{C}$ is the normalization constant and the valid radial coordinate range is $\rho_{\circ}=\beta/\alpha\leq r<\infty$, where $\rho_{\circ}$ describes the origin of the exterior radial coordinate, the region of interest of the current study, within which the radial wave function is valid and normalized. Hence, one would enforce the continuity conditions on the radial wave function at $\rho_{\circ}=\beta/\alpha$ so that
\begin{equation}
    R_{int.}(\rho)|_{\rho=\rho_{\circ}}=R_{ext.}(r)|_{r=\rho_{\circ}};\quad R'_{int.}(\rho)|_{\rho=\rho_{\circ}}=R'_{ext.}(r)|_{r=\rho_{\circ}}, \label{2.60}
\end{equation}
where 
\begin{equation}
R_{int.}(\rho)=\frac{U\left( \rho\right)}{\sqrt{\rho}} =\mathcal{C}\,\rho^{|\tilde{m}|}e^{-\tilde{\Omega}%
\rho^{2}/2}\,_{1}F_{1}\left( \frac{1}{2}+\frac{|\tilde{m}|}{2}-\frac{\mathcal{%
\tilde{E}}^{2}}{4\tilde{\Omega}},|\tilde{m}|+1,\tilde{\Omega}\rho^{2}\right),
\label{2.61}
\end{equation}%
with $\rho\in[0,\rho_{\circ}=\beta/\alpha]$, $R_{int.}(\rho)$,  (\ref{2.61}), and $R_{ext.}(r)$, (\ref{2.6}), are the radial wave functions in the internal and external (without the internal structure) cosmic string structures, respectively. Obviously, the continuity conditions in (\ref{2.60}) are clearly satisfied.}

The result (\ref{2.6}) is, moreover, in exact agreement with that reported in Eq.(12) as $R^{(1)}(r)$ in \cite{1.19}.  Of course, there is another linearly independent solution given as $R^{(2)}(r)\sim r^{-|\tilde{m}|}\,e^{-\tilde{\Omega}%
r^{2}/2}$ in  \cite{1.19}. However,   $R^{(2)}(r)$ should be dismissed as it is not finite at $r=0$ when $\beta=0$ (i.e., for the special case of a static non-spinning cosmic string, which is a special case of the more general spinning string when $\beta\neq0$ considered here). One should also be aware that the asymptotic behaviors of the radial function $R(r)=R^{(1)}(r)$  are readily given by $r^{|\tilde{m}|}$ as $r\rightarrow0$, for $\beta=0$, and $e^{-\tilde{\Omega}r^{2}/2}$ as $r\rightarrow\infty$. Therefore, the finiteness of the radial function $R(r)$ in (\ref{2.6}) is secured by the exponential term, while the square integrability is secured by the requirement that the confluent power series be truncated to a polynomial of order $n_{r}\geq 0$. This is done by the requirement that%
\begin{equation}
\frac{1}{2}+\frac{|\tilde{m}|}{2}-\frac{\mathcal{\tilde{E}}^{2}}{4\tilde{%
\Omega}}=-n_{r}\Longrightarrow \mathcal{\tilde{E}}^{2}=2\tilde{\Omega}\left(
2n_{r}+|\tilde{m}|+1\right) .  \label{2.7}
\end{equation}%
In the appendix, we show a detailed power series solution and show exactly how to obtain such an exact result.

We may now use (\ref{2.4}) and (\ref{2.5}) in (\ref{2.7}) to obtain%
\begin{equation}
E^{2}+\frac{eB}{\alpha }\left( \beta E+m\right) -\frac{2\tilde{\Omega}}{%
\alpha }\left\vert \beta E+m\right\vert -\mathcal{G}_{n_{r}}=0\,,
\label{2.8}
\end{equation}%
where%
\begin{equation}
\mathcal{G}_{n_{r}}=k^{2}+m_{\circ }^{2}+2\Omega +2\tilde{\Omega}\left(
2n_{r}+1\right) .  \label{2.9}
\end{equation}

At this point, however, one should observe that our result in (\ref{2.8}) is in exact agreement (when our $\Omega =0$ in our $\mathcal{F}_{r}=\Omega r$) with that reported by Cunha et al. \cite{1.19} (should the typo in their Eq. (15) be corrected, i.e., last term on the LHS should be removed). One should also be aware of the fact that $E=E_{\pm }=\pm |E|$ and the magnetic quantum number $m=m_{\pm }=\pm |m|$. Notably, for a static cosmic string $%
\beta =0$, the Landau levels for the KG-oscillators energies are given by%
\begin{equation}
E_{\pm }=\pm \sqrt{\mathcal{G}_{n_{r}}+\frac{2\tilde{\Omega}}{\alpha }%
\left\vert m\right\vert -\frac{eB}{\alpha }m},  \label{2.9.1}
\end{equation}%
which is in exact accord with that reported in Eq. (23) of Medeiros et al. \cite{1.24}. Moreover, one should observe that the accidental degeneracies associated with $m=m_{+}=\left\vert m\right\vert $ are\ now lifted/dismissed by the KG-oscillator's frequency $\Omega $ in $\tilde{\Omega}^{2}=\tilde{B}%
^{2}+\Omega ^{2}$. However, the spinning/rotation of the cosmic string manifestly introduces an additional complexity to the problem at hand. Equation (\ref{2.8}) should therefore be dealt with in an orderly manner (i.e., case by case). We do this in the sequel.

\subsection{The simplistic case $m=0$}

For the case where the magnetic quantum number $m=0$ (i.e., $S$-states), Eq. (\ref{2.8}) would
imply%
\begin{equation}
E^{2}+\frac{eB}{\alpha }\beta E-\frac{2\tilde{\Omega}}{\alpha }\beta
\left\vert E\right\vert -\mathcal{G}_{n_{r}}=0\Longrightarrow E_{\pm
}^{2}+E_{\pm }\left( \frac{\beta }{\alpha }\right) \left[ eB\mp 2\tilde{%
\Omega}\right] -\mathcal{G}_{n_{r}}=0.  \label{2.10}
\end{equation}%
This would in turn yield%
\begin{equation}
E_{\pm }=\frac{-\beta \left[ eB\mp 2\tilde{\Omega}\right] \pm \sqrt{\beta
^{2}\left[ eB\mp 2\tilde{\Omega}\right] ^{2}+4\alpha ^{2}\mathcal{G}_{n_{r}}}%
}{2\alpha }.  \label{2.11}
\end{equation}%
This result would necessarily mean that the particles' energies $E_{+}$ are given by%
\begin{equation}
E_{+}=\frac{-\beta \left[ eB-2\tilde{\Omega}\right] +\sqrt{\beta ^{2}\left[
eB-2\tilde{\Omega}\right] ^{2}+4\alpha ^{2}\mathcal{G}_{n_{r}}}}{2\alpha },
\label{2.12}
\end{equation}%
whereas, the antiparticles' energies are given by%
\begin{equation}
E_{-}=\frac{-\beta \left[ eB+2\tilde{\Omega}\right] -\sqrt{\beta ^{2}\left[
eB+2\tilde{\Omega}\right] ^{2}+4\alpha ^{2}\mathcal{G}_{n_{r}}}}{2\alpha }.
\label{2.13}
\end{equation}%
At this point, one may argue that this case is perhaps the most proper one that allows us to build a clear picture of the effects of a spinning/rotating cosmic string on the spectroscopic structure of the $S$-states of the KG-oscillators. Apparently, the results in (\ref{2.12}) and (\ref{2.13}) show that the symmetry of the positive and negative energies about the $E=0$ is broken as a result of the cosmic string spinning $\beta \neq 0
$. However, such symmetry can be recovered for a static cosmic string $\beta =0$. It is, moreover, clear that for $\Omega=0\rightarrow \tilde{\Omega}=eB/2$ and the KG-particles' and the KG-antiparticles' $S$-states, respectively, read%
\begin{equation}
E_+=+\sqrt{\mathcal{G}_{n_{r}}}, \, \, E_{-}=\frac{-\beta  eB\ -\sqrt{\beta ^{2}
e^2B^{2}+\alpha ^{2}\mathcal{G}_{n_{r}}}}{\alpha }.
\label{2.13.1}
\end{equation}%
\begin{figure}[ht!]  
\centering
\includegraphics[width=0.3\textwidth]{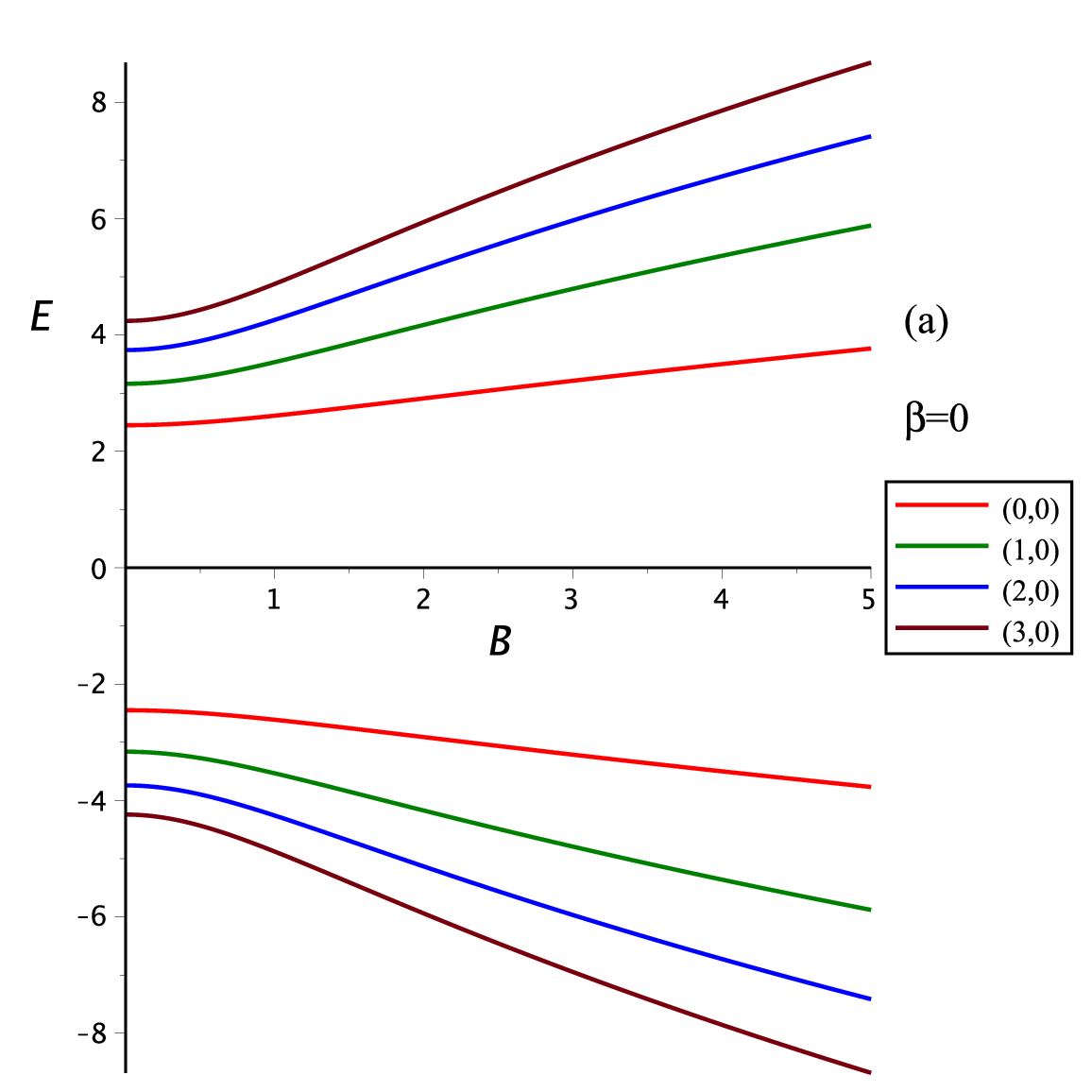}
\includegraphics[width=0.3\textwidth]{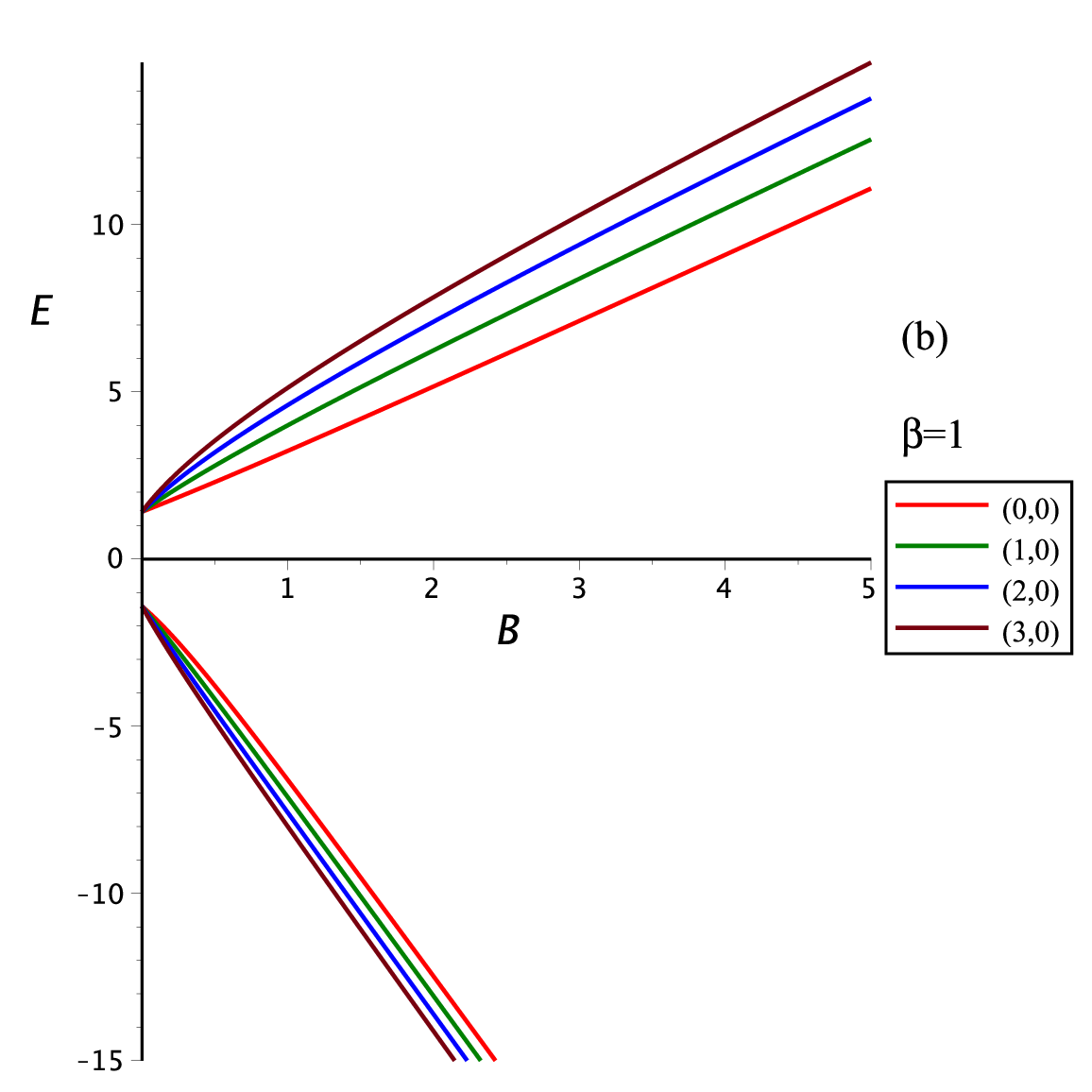}
\includegraphics[width=0.3\textwidth]{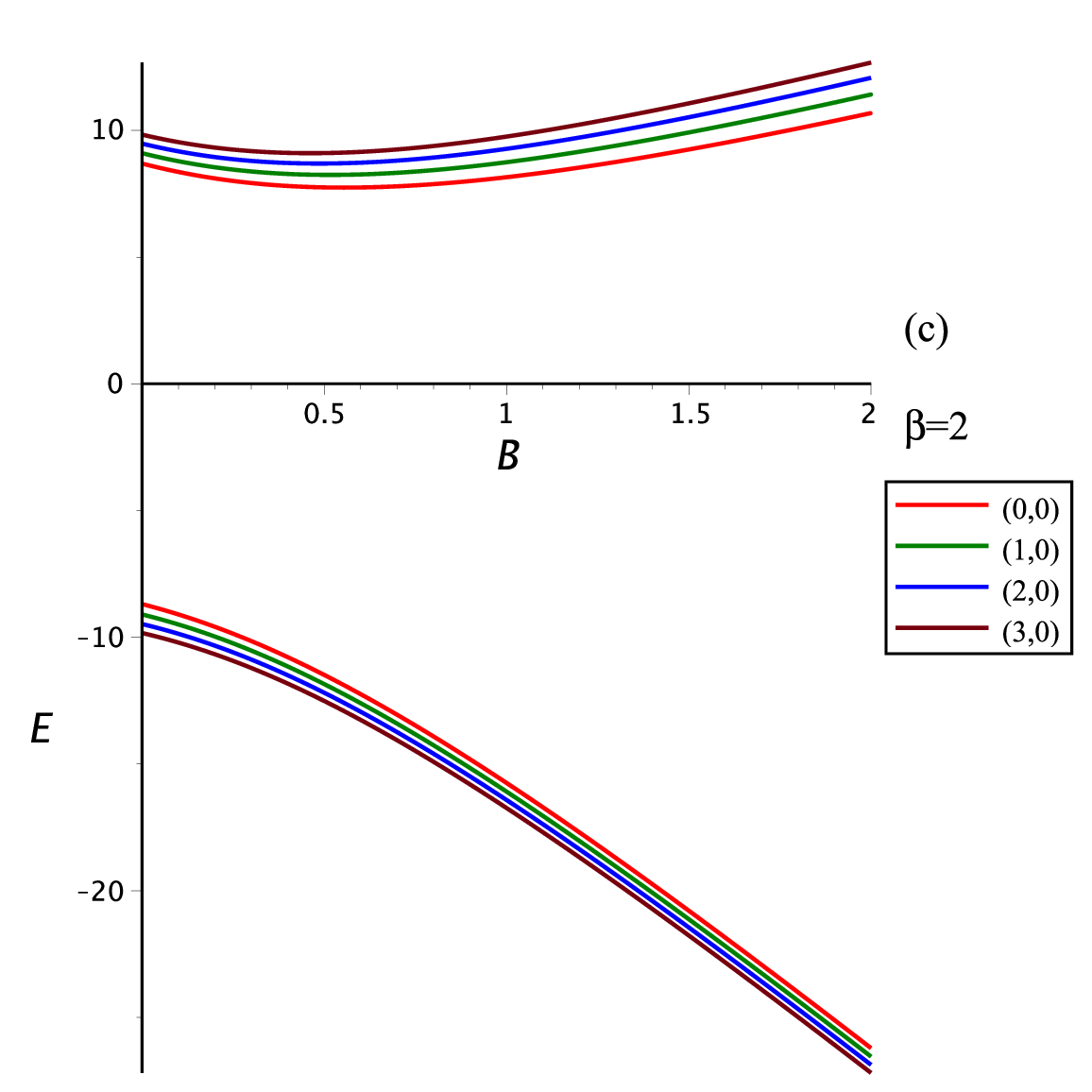}
\caption{\small 
{ The energy levels against the magnetic field strength $B$
(for the magnetic quantum number $m=0$ and $n_{r}=0,1,2,3$) given by Eq. (%
\ref{2.10}), at $\alpha =0.5$, and $m_{\circ }=1=e=k=\Omega $, so that
Fig.1(a) for $\beta =0$ ( no spinning of the cosmic string), 1(b) for $\beta
=1$, and 1(c) for $\beta =2$.}}
\label{fig1}
\end{figure}%
\begin{figure}[ht!]  
\centering
\includegraphics[width=0.3\textwidth]{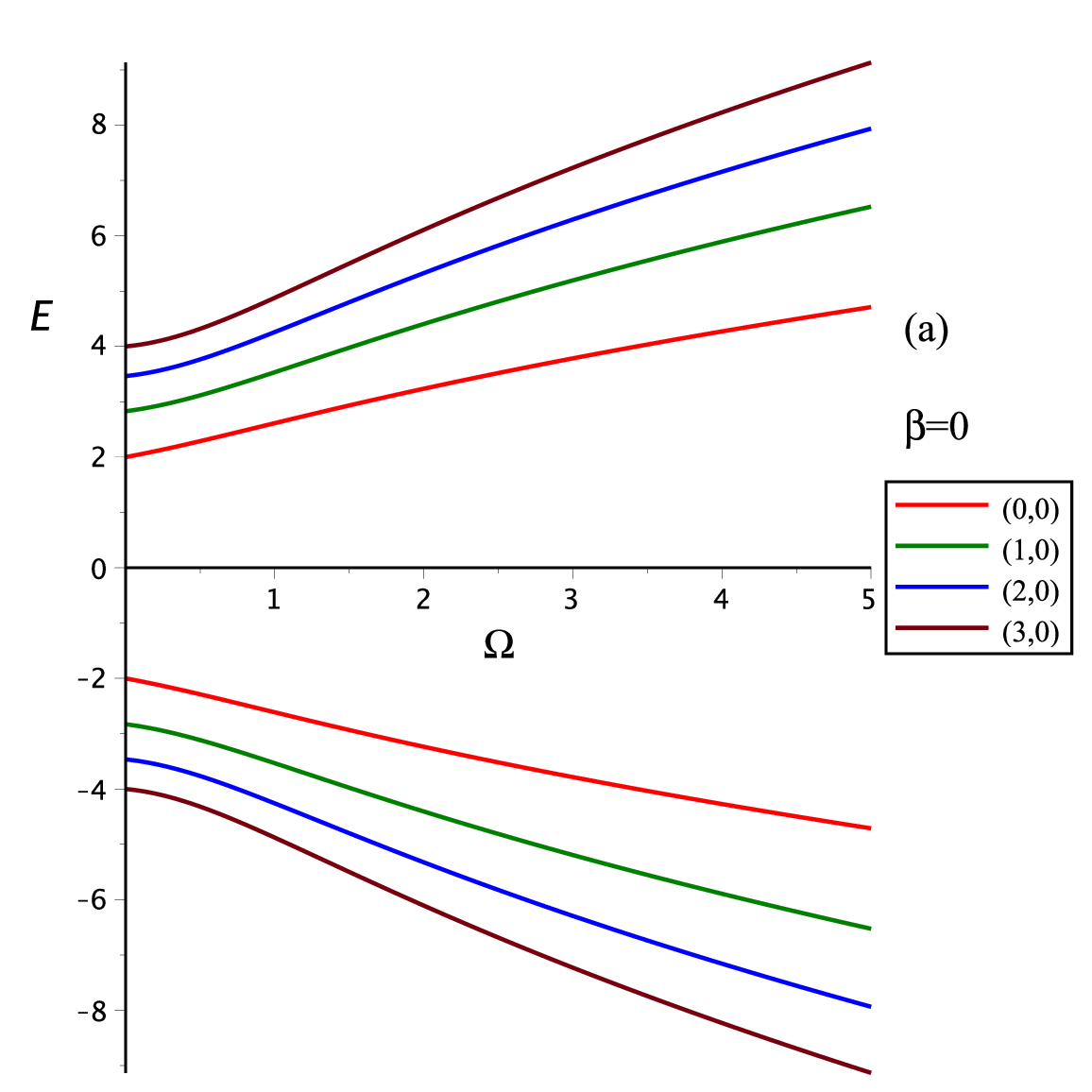}
\includegraphics[width=0.3\textwidth]{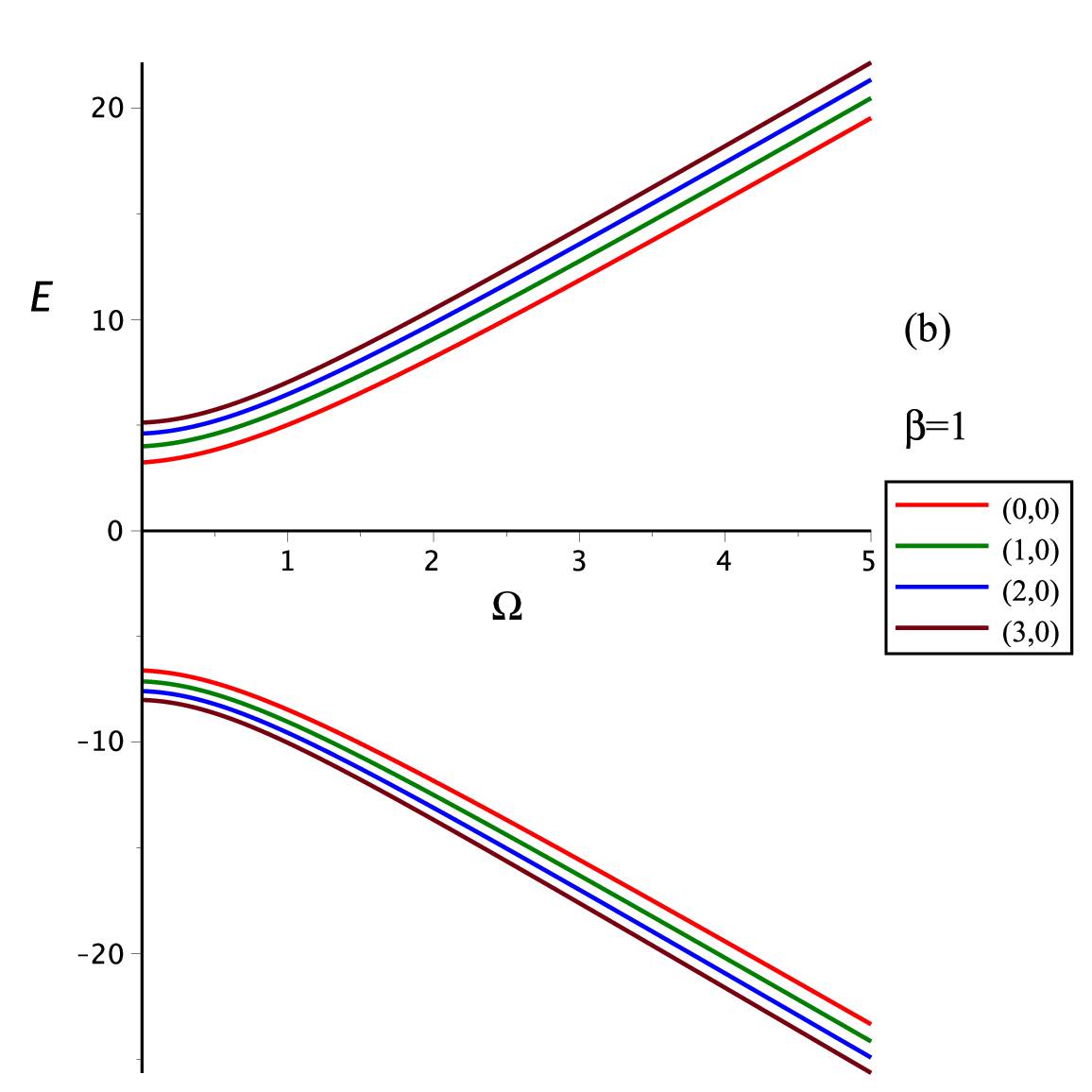}
\includegraphics[width=0.3\textwidth]{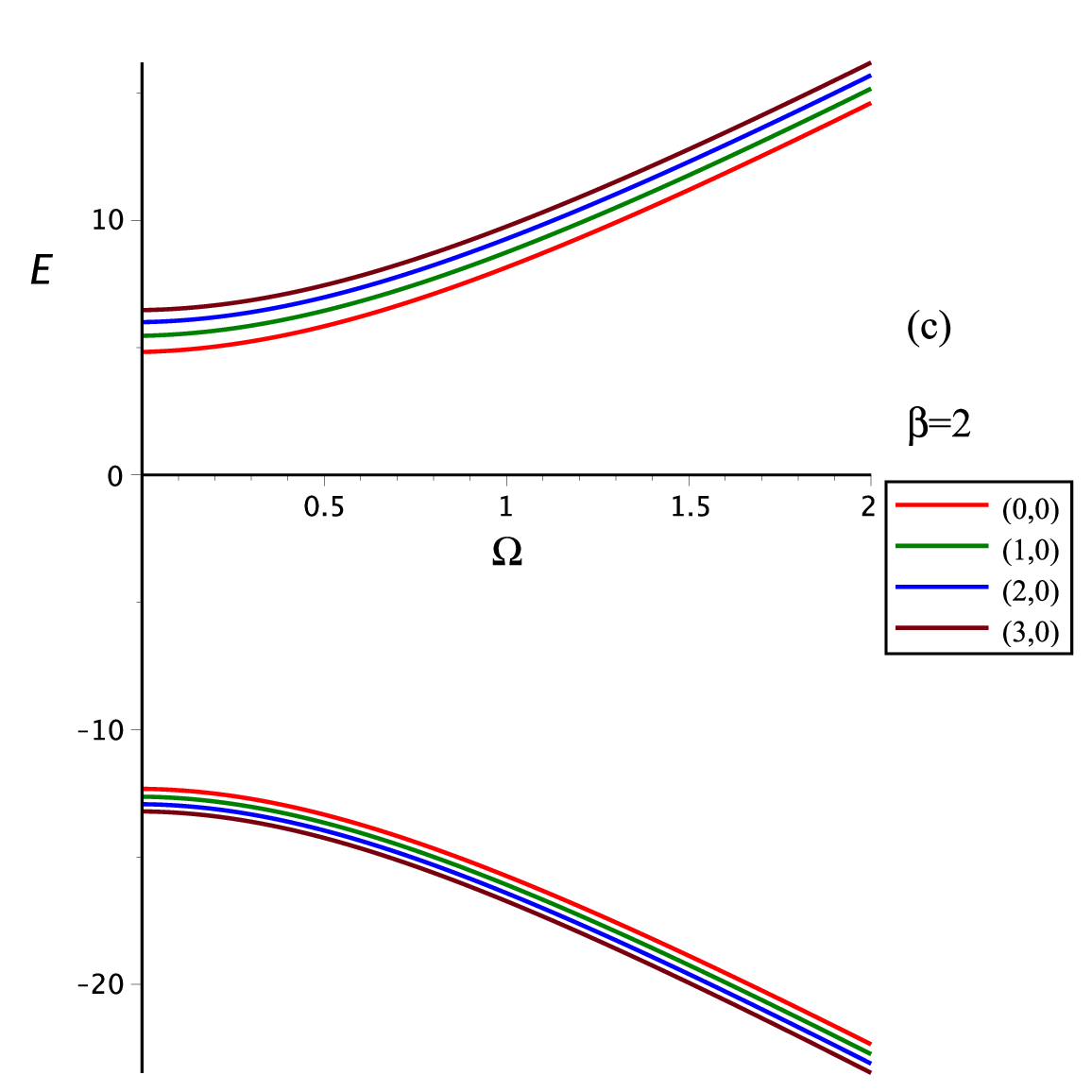}
\caption{\small 
{ The energy levels against the KG-oscillators' frequency $%
\Omega $  (for $m=0$ and $n_{r}=0,1,2,3$) given by Eq. (\ref{2.10}), at $%
\alpha =0.5$, and $m_{\circ }=1=e=k=B$, so that 2(a) for $\beta =0$, 2(b)
for $\beta =1$, and 2(c) for $\beta =2$.}}
\label{fig2}
\end{figure}%
This result obviously suggests that the KG-particles' $S$-states are not infected by neither the spinning parameter $\beta$ nor by the wedge parameter $\alpha$, whereas the KG-antiparticles' $S$-states are indeed infected. This would, in turn, suggest that the co-existence of a spinning cosmic string and a magnetic field  has eliminated/cancelled  the effects of the wedge parameter $\alpha$ on the $S$-states of the KG-particles., but not on the antiparticles. Notably, however, switching off the magnetic field, $B=0$, and setting $\Omega\neq0$, one obtains%
\begin{equation}
E_{\pm}=\frac{\pm \beta\Omega \pm\sqrt{\beta ^{2}
\Omega^{2}+\alpha ^{2}\mathcal{G}_{n_{r}}}}{\alpha },
\label{2.13.2}
\end{equation}%
which clearly documents symmetrization of the KG-oscillators' energies about $E=0$ for particles and antiparticles. Yet, the effects of the wedge parameter $\alpha$ and the spinning parameter $\beta$ have their fingerprints on the spectrum of the KG-oscillators. Interestingly, the comparison between (\ref{2.13.1}) and (\ref{2.13.2}) indicates that the co-existence of both a spinning cosmic string and an external magnetic field would eliminate/cancel the wedge parameter effect (a cosmic string byproduct) for particles' energies, and would, moreover, break the symmetry of the particles' and antiparticles' energies about $E=0$.

In Figures 1 and 2, we show the corresponding energies, for the magnetic quantum number $m=0$ and $n_{r}=0,1,2,3$, given by Eq. (\ref{2.10}) at $%
\alpha =0.5$, and $m_{\circ }=1=e=k$. Where, in Fig.1 we show the energies
for $\Omega =1$ against the magnetic field strength $B$, so that Fig.1(a) for $\beta =0$ ( no spinning of the cosmic string), 1(b) for $\beta =1$, and 1(c) for $\beta =2$. Whereas, in Fig.2, we show the energies for $B=1$ against the KG-oscillator's frequency $\Omega $, so that 2(a) for $\beta =0$, 2(b) for $\beta =1$, and 2(c) for $\beta =2$. A common trend in the figures suggests that clustering of the energy levels for KG-oscillators (both particles and anti-particles) is eminent for $\beta >>1$.  This would, in turn, indicate that there is no distinction between energy levels at such values of $\beta $.

\subsection{The energy sets for the case where $E=E_{\pm }$ and $m=m_{\pm }$}

The case where $E=E_{\pm }=\pm \left\vert E\right\vert $ and $m=m_{\pm }=\pm
\left\vert m\right\vert $ would specifically identify a set of particles, $%
E=E_{+}=+\left\vert E\right\vert $, associated with the magnetic number $%
m=m_{+}=+\left\vert m\right\vert $, and a set of anti-particles, $%
E=E_{-}=-\left\vert E\right\vert $, associated with $m=m_{-}=-\left\vert
m\right\vert $, each set at a time. In this case, our result in (\ref{2.8})
would imply%
\begin{equation}
E_{\pm }^{2}+\frac{eB}{\alpha }m_{\pm }\left( \frac{\beta E_{\pm }}{m_{\pm }}%
+1\right) -\frac{2\tilde{\Omega}}{\alpha }\left\vert m\right\vert \left\vert 
\frac{\beta E_{\pm }}{m_{\pm }}+1\right\vert -\mathcal{G}_{n_{r}}=0.\,
\label{2.14}
\end{equation}%
One should notice that this result holds true if and only if the KG-particles, $E=E_{+}$, are accompanied by $m=m_{+}$ and the KG-anti-particles, $E=E_{-}$, are accompanied by $m=m_{-}$. Moreover, this equation would immediately, with $\left\vert m\right\vert =\pm m_{\pm }$, imply 
\begin{equation}
E_{\pm }^{2}+\left( \frac{\beta E_{\pm }}{m_{\pm }}+1\right) \left( \frac{%
m_{\pm }}{\alpha }\right) \left[ eB\mp 2\tilde{\Omega}\right] -\mathcal{G}%
_{n_{r}}=0.  \label{2.15}
\end{equation}%
One may simplify this equation to read 
\begin{equation}
E_{\pm }^{2}+\frac{\beta E_{\pm }}{\alpha }\left[ eB\mp 2\tilde{\Omega}%
\right] -\mathcal{G}_{n_{r},m_{\pm }}=0\,;\;\mathcal{G}_{n_{r},m_{\pm }}=%
\mathcal{G}_{n_{r}}-\frac{m_{\pm }}{\alpha }\left[ eB\mp 2\tilde{\Omega}%
\right] .  \label{2.16}
\end{equation}%
The exact translation of which should yield 
\begin{equation}
E_{+}^{2}+\frac{\beta E_{+}}{\alpha }\left[ eB-2\tilde{\Omega}\right] -%
\mathcal{G}_{n_{r},m_{+}}=0\,;\;\mathcal{G}_{n_{r},m_{+}}=\mathcal{G}%
_{n_{r}}-\frac{\left\vert m\right\vert }{\alpha }\left[ eB-2\tilde{\Omega}%
\right] ,  \label{2.17}
\end{equation}%
and%
\begin{equation}
E_{-}^{2}+\frac{\beta E_{-}}{\alpha }\left[ eB+2\tilde{\Omega}\right] -%
\mathcal{G}_{n_{r},m_{-}}=0;\;\mathcal{G}_{n_{r},m_{-}}=\mathcal{G}_{n_{r}}+%
\frac{\left\vert m\right\vert }{\alpha }\left[ eB+2\tilde{\Omega}\right] .
\label{2.18}
\end{equation}%
Consequently, one may conclude that%
\begin{equation}
E_{\pm }=\frac{-\beta \left[ eB\mp 2\tilde{\Omega}\right] \pm \sqrt{\beta
^{2}\left[ eB\mp 2\tilde{\Omega}\right] ^{2}+4\alpha ^{2}\mathcal{G}%
_{n_{r},m_{\pm }}}}{2\alpha }.  \label{2.19}
\end{equation}%
This result would, in turn, retrieve the one reported by Cunha et al. \cite{1.19} (i.e., their Eq. (16)) for our $\Omega =0$ and our $\tilde{\Omega}=eB/2$), provided that the two sets $E_{+}$ with $m_{+}
$ and $E_{-}$ with $m_{-}$ are respected. That is, Cunha et al. \cite{1.19} have used the term $\left( \left\vert \ell \right\vert -\ell \right) /\ell $  in their Eq. (16) so that $\left( \left\vert \ell \right\vert -\ell \right)
/\ell =0$ for $\ell =\ell _{+}=+\left\vert \ell \right\vert $ ($m_{+}$ in the current proposal) and $\left( \left\vert \ell \right\vert -\ell \right)
/\ell =-2$ for $\ell =\ell _{-}=-\left\vert \ell \right\vert $ ($m_{-}$ in
the current proposal). Moreover, their term $\left( \left\vert \ell
\right\vert -\ell \right) =0$ for $\ell =\ell _{+}=+\left\vert \ell
\right\vert $ and $\left( \left\vert \ell \right\vert -\ell \right)
=2\left\vert \ell \right\vert $ for $\ell =\ell _{-}=-\left\vert \ell
\right\vert $. The mapping between our results and the result of Cunha et al. \cite{1.19} is clear, therefore.

Obviously, however, for $\Omega =0$ (i.e., the case considered by Cunha et al. \cite{1.19}), one gets $\tilde{\Omega}=eB/2$  to yield
that the KG-particles' and KG-antiparticles' energies are given, respectively, by%
\begin{equation}
E_{+}=+\sqrt{\mathcal{G}_{n_{r}}},\text{ \ }E_{-}=-\frac{eB\beta }{\alpha }-%
\frac{1}{\alpha }\sqrt{e^{2}B^{2}\beta ^{2}+\alpha ^{2}\mathcal{G}_{n_{r}}}.
\label{2.21}
\end{equation}%
\begin{figure}[ht!]  
\centering
\includegraphics[width=0.3\textwidth]{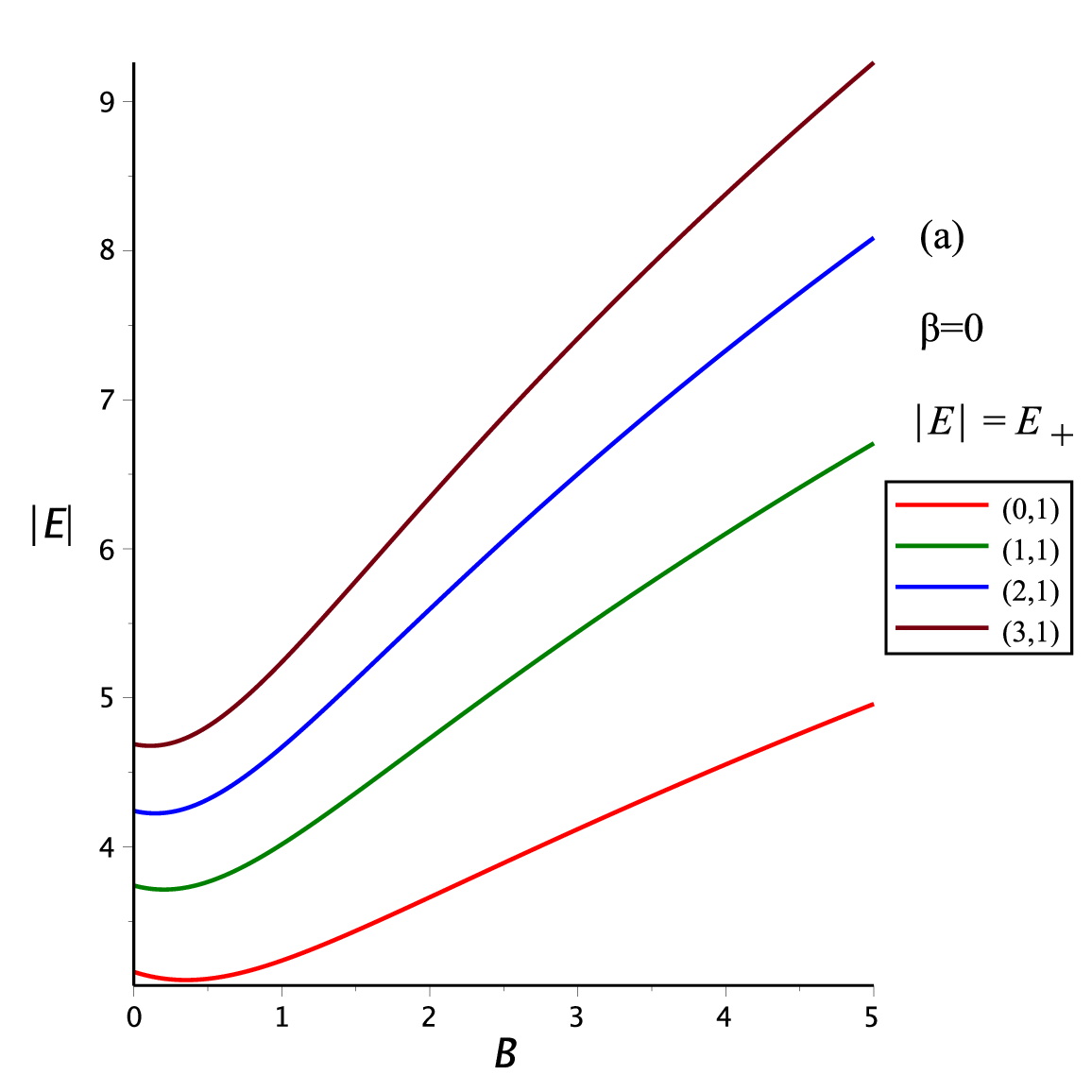}
\includegraphics[width=0.3\textwidth]{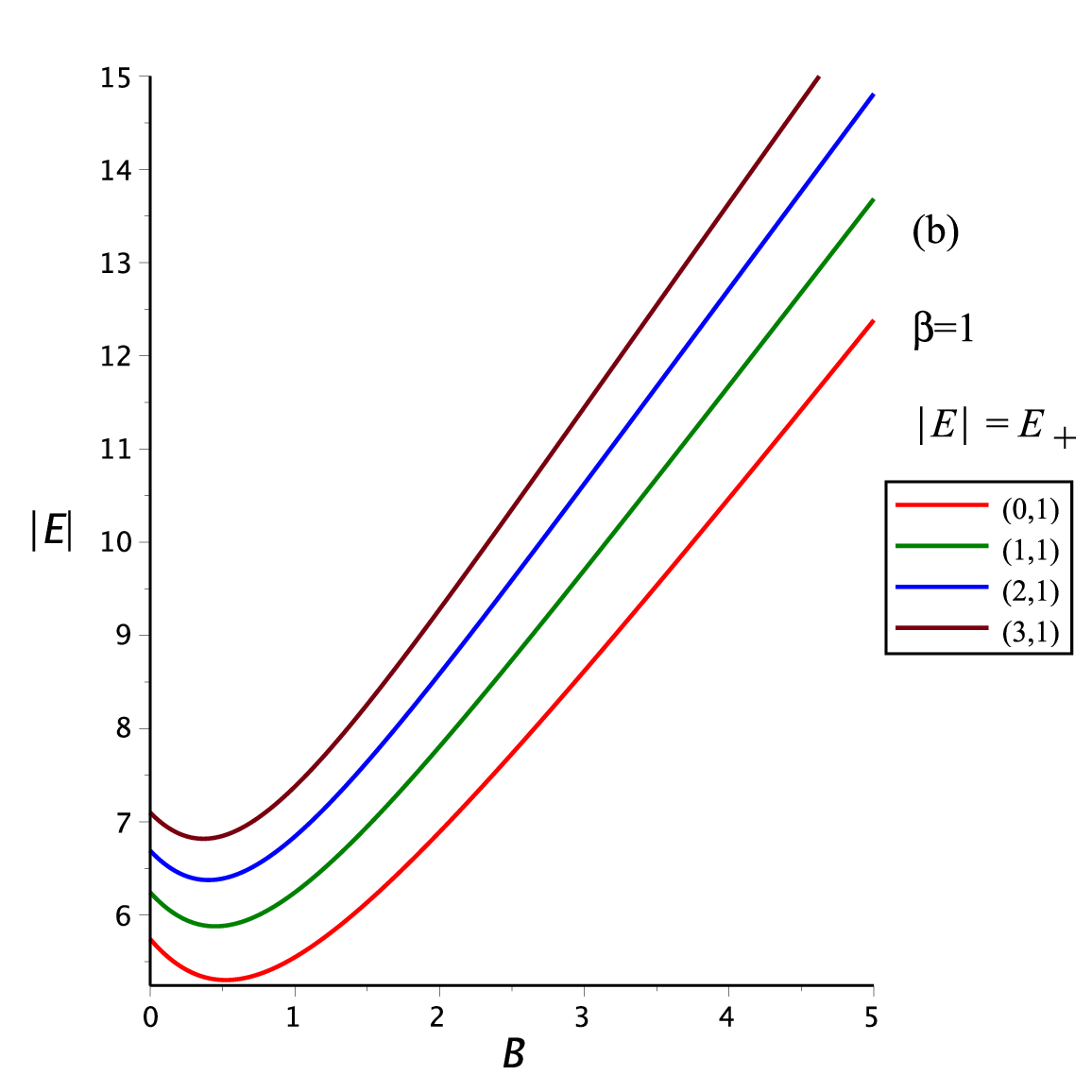}
\includegraphics[width=0.3\textwidth]{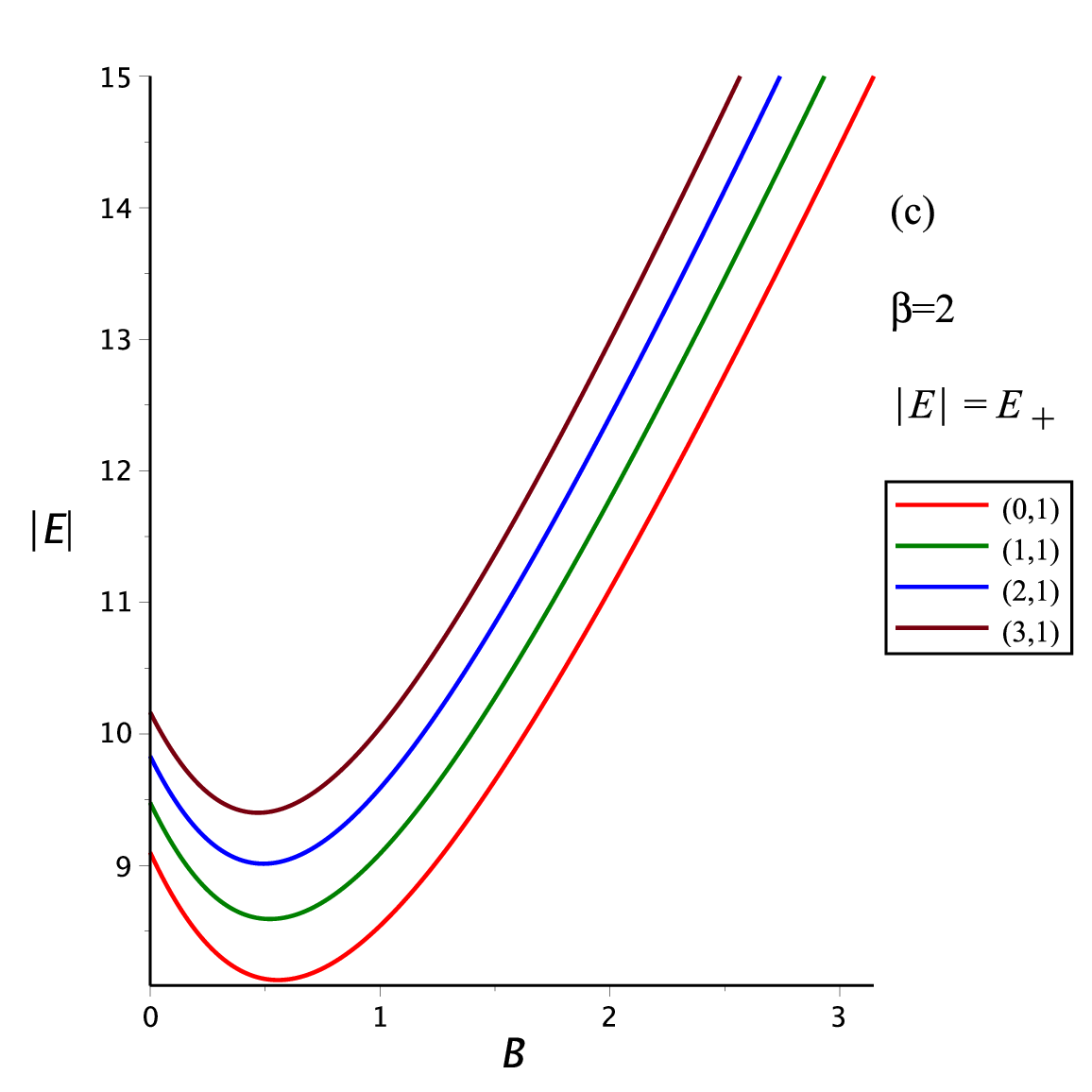}
\includegraphics[width=0.3\textwidth]{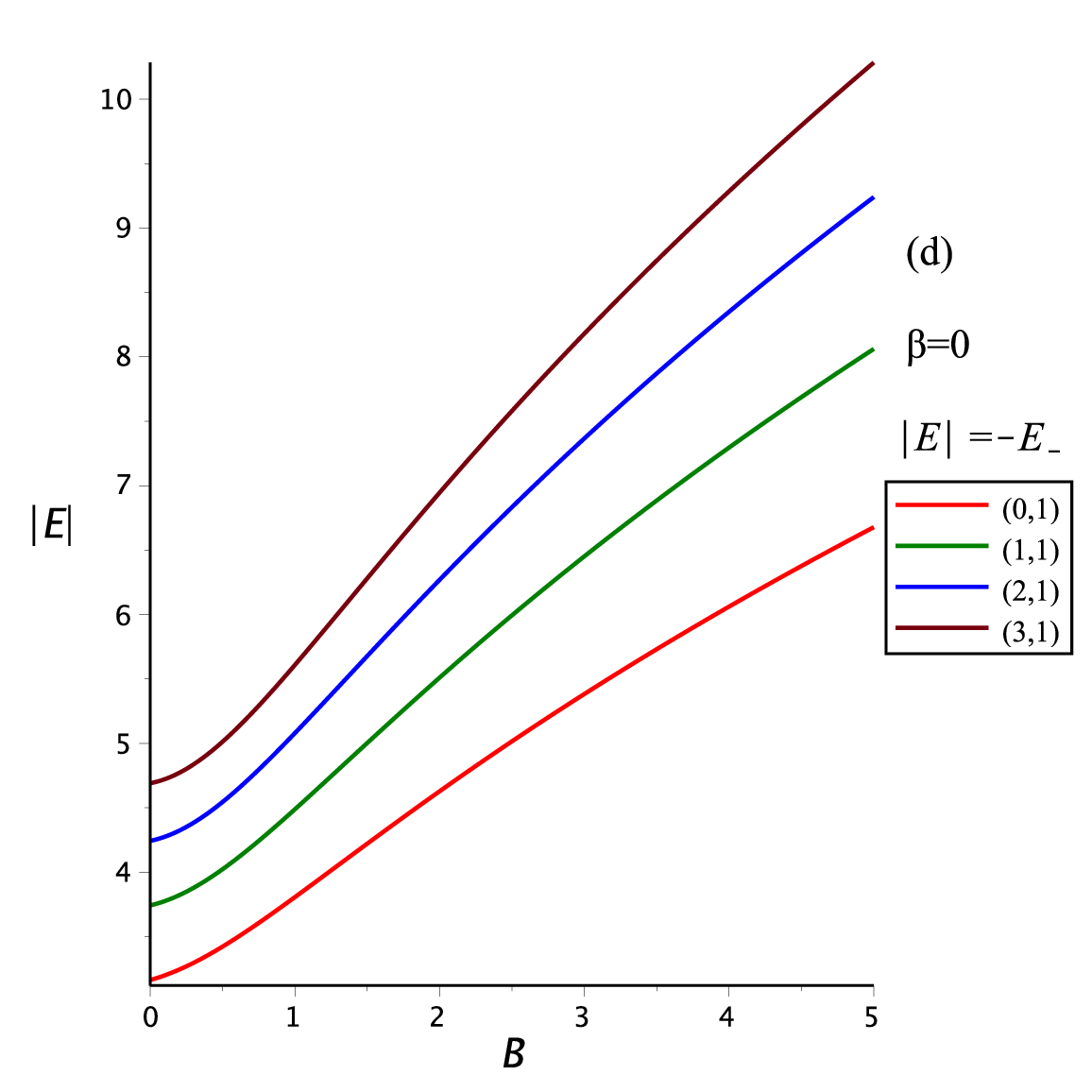}
\includegraphics[width=0.3\textwidth]{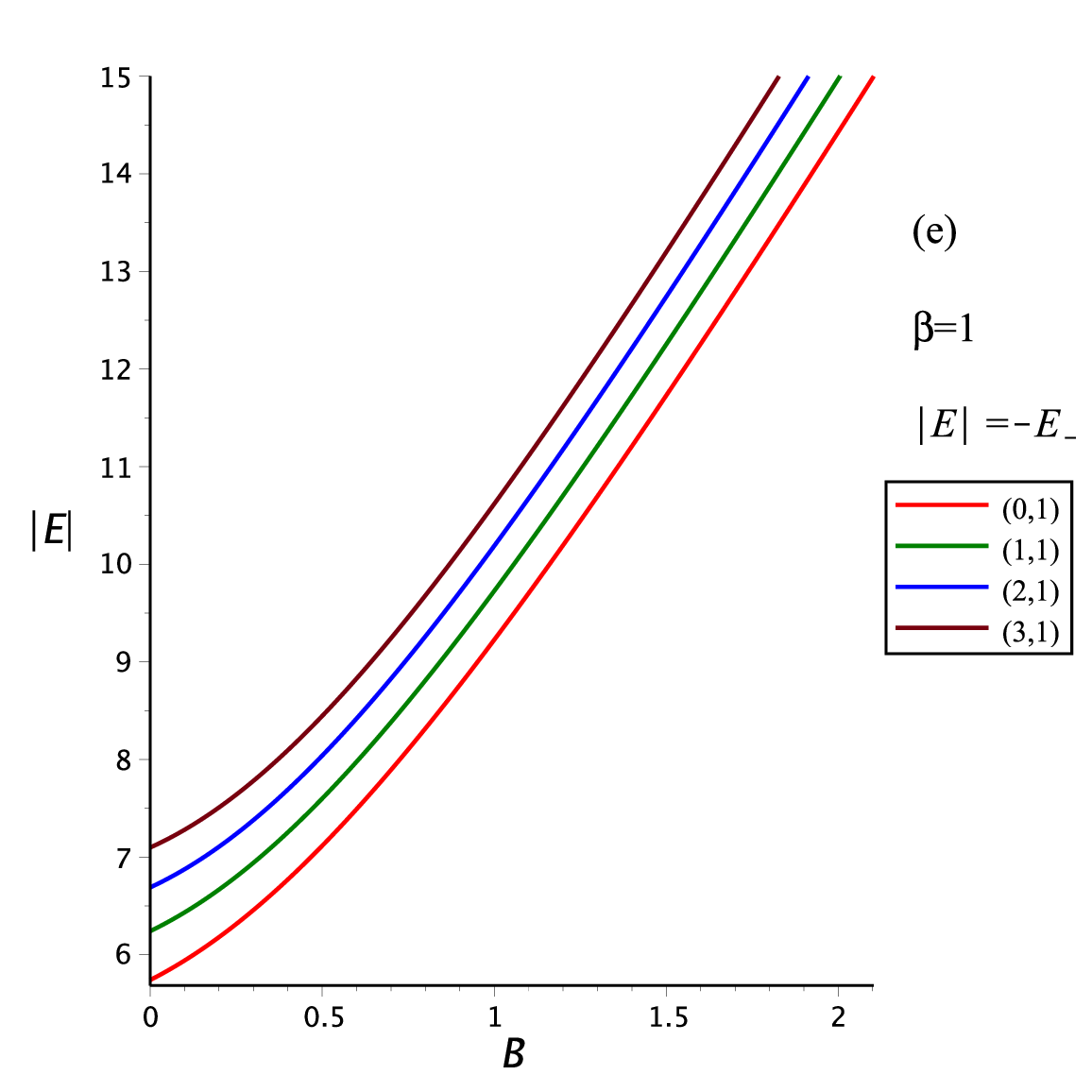}
\includegraphics[width=0.3\textwidth]{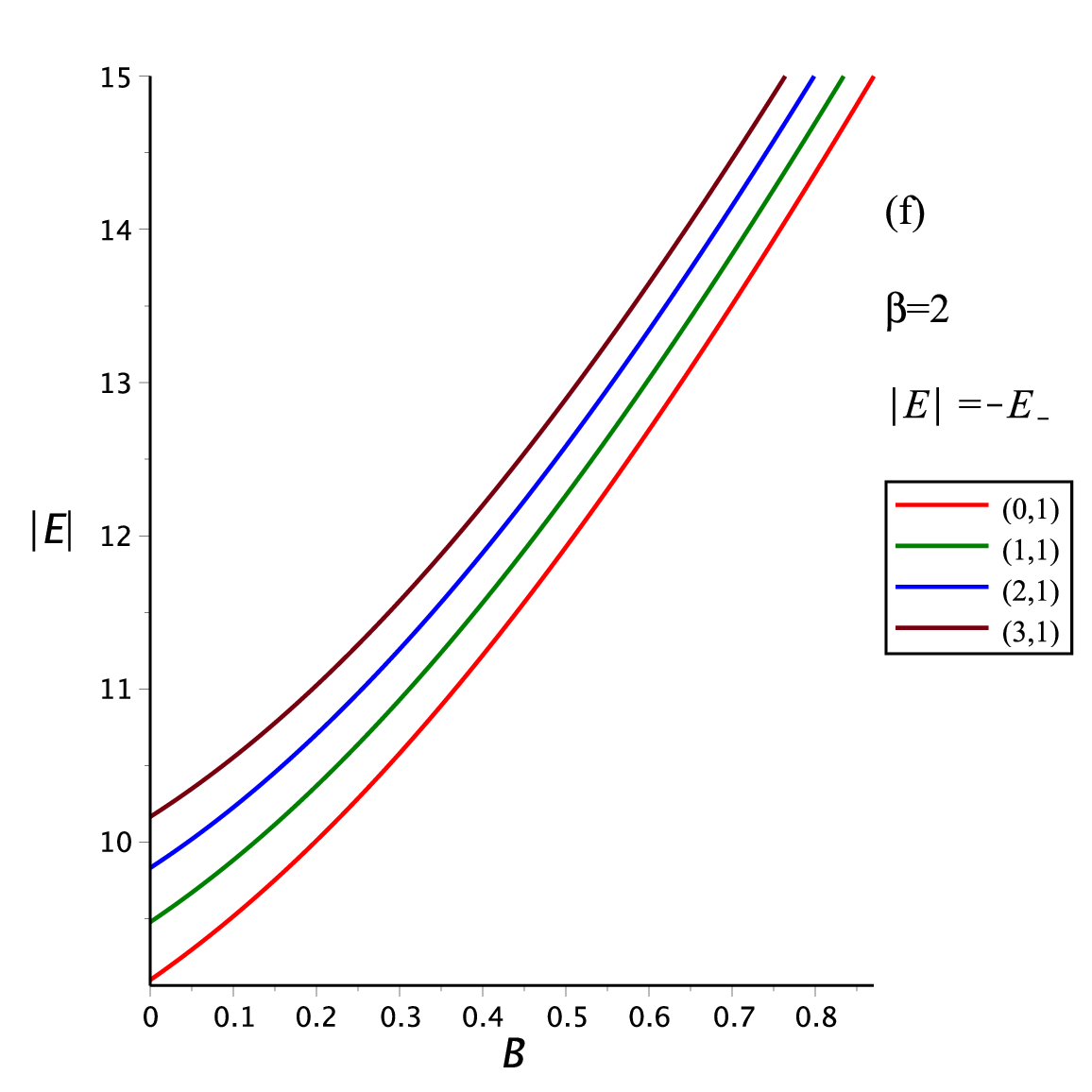}
\caption{\small 
{ The energy levels (\ref{2.19}) against the magnetic field
strength $B$ (for $m=1\neq 0$ and $n_{r}=0,1,2,3$ at $\alpha =0.5$, and $%
m_{\circ }=1=e=k=\Omega $) so that the particles' energies are given in 3(a)
for $\beta =0$, 3(b) for $\beta =1$, and 3(c) for $\beta =2$. The
anti-particles' energies are given in 3(d) for $\beta =0$, 3(e) for $\beta =1
$, and 3(f) for $\beta =2$.}}
\label{fig3}
\end{figure}%
This result clearly suggests that the the energies of KG-particles, $E=E_{+}$, accompanied by $m=m_{+}$ under such settings (i.e., $\Omega =0$,) are not infected by the spinning/rotation of the cosmic string, whereas
KG-antiparticles, $E=E_{-}$, accompanied by $m=m_{-}$, are. Yet, interestingly, we again observe that the spinning string has removed/eliminated the effect of the wedge parameter $\alpha $ (a byproduct of the cosmic string itself) on the KG-particles energies, whereas the wedge parameter's effect is obvious on the KG-antiparticles. 

Notably, moreover, switching off the magnetic field (i.e., $B=0$ ) would reduce our result in (\ref{2.19}) to read%
\begin{equation}
E_{\pm }=\pm \frac{\beta \Omega }{\alpha }\pm \frac{1}{\alpha }\sqrt{\beta
^{2}\Omega ^{2}+\alpha ^{2}\mathcal{G}_{n_{r},m_{\pm }}};\;\mathcal{G}%
_{n_{r},m_{\pm }}=\mathcal{G}_{n_{r}}+\frac{2\Omega \left\vert m\right\vert 
}{\alpha },  \label{2.21.1}
\end{equation}%
which clearly and interestingly indicates that the energy levels are symmetric about $E=0$, and  the energies of the KG-particles are mirror reflection, about $E=0$, of those of the antiparticles and vice versa. Consequently,  the simultaneous presence of a spinning cosmic string along with an external magnetic field would break such a symmetry as documented in (\ref{2.19}) and (\ref{2.21}) . 
\begin{figure}[ht!]  
\centering
\includegraphics[width=0.3\textwidth]{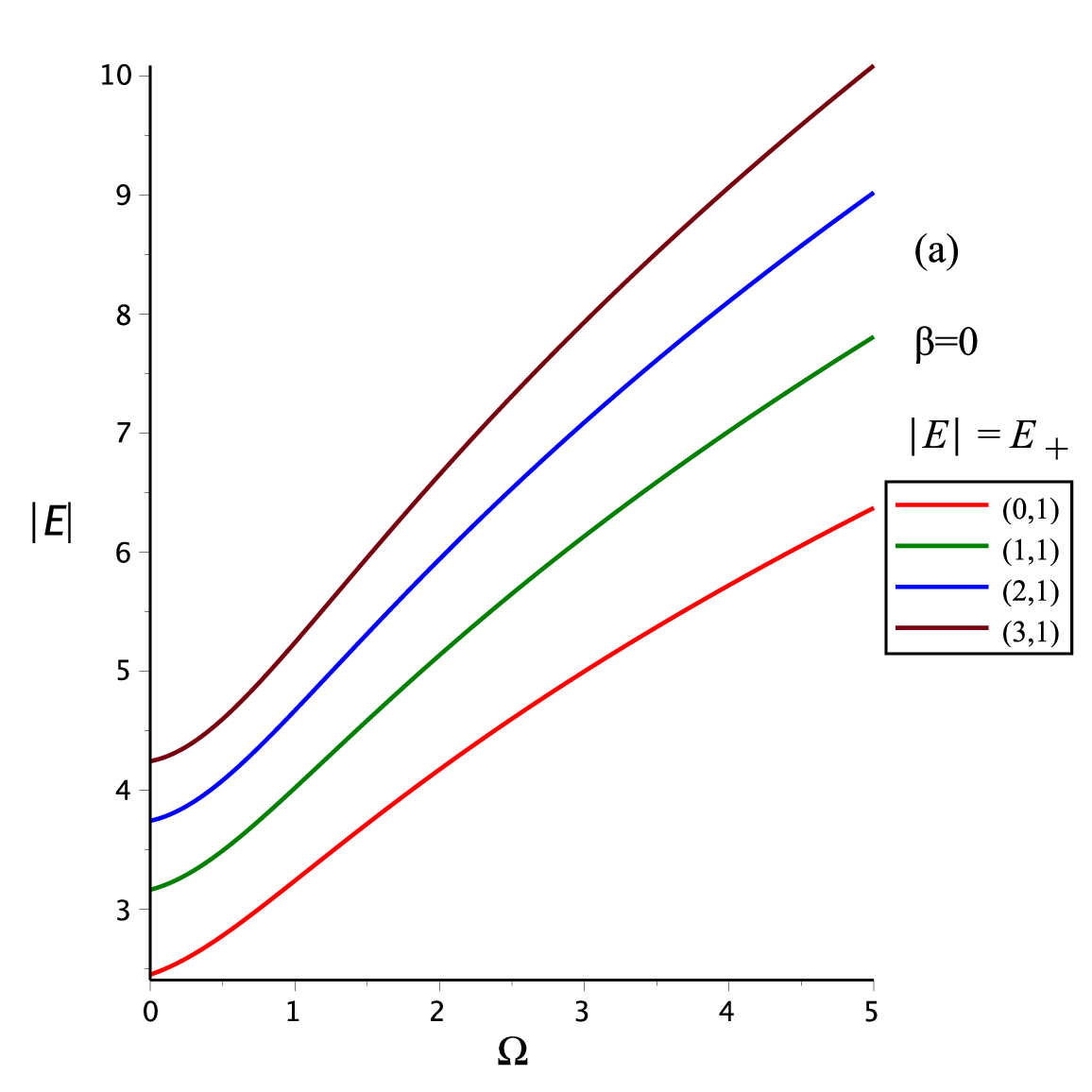}
\includegraphics[width=0.3\textwidth]{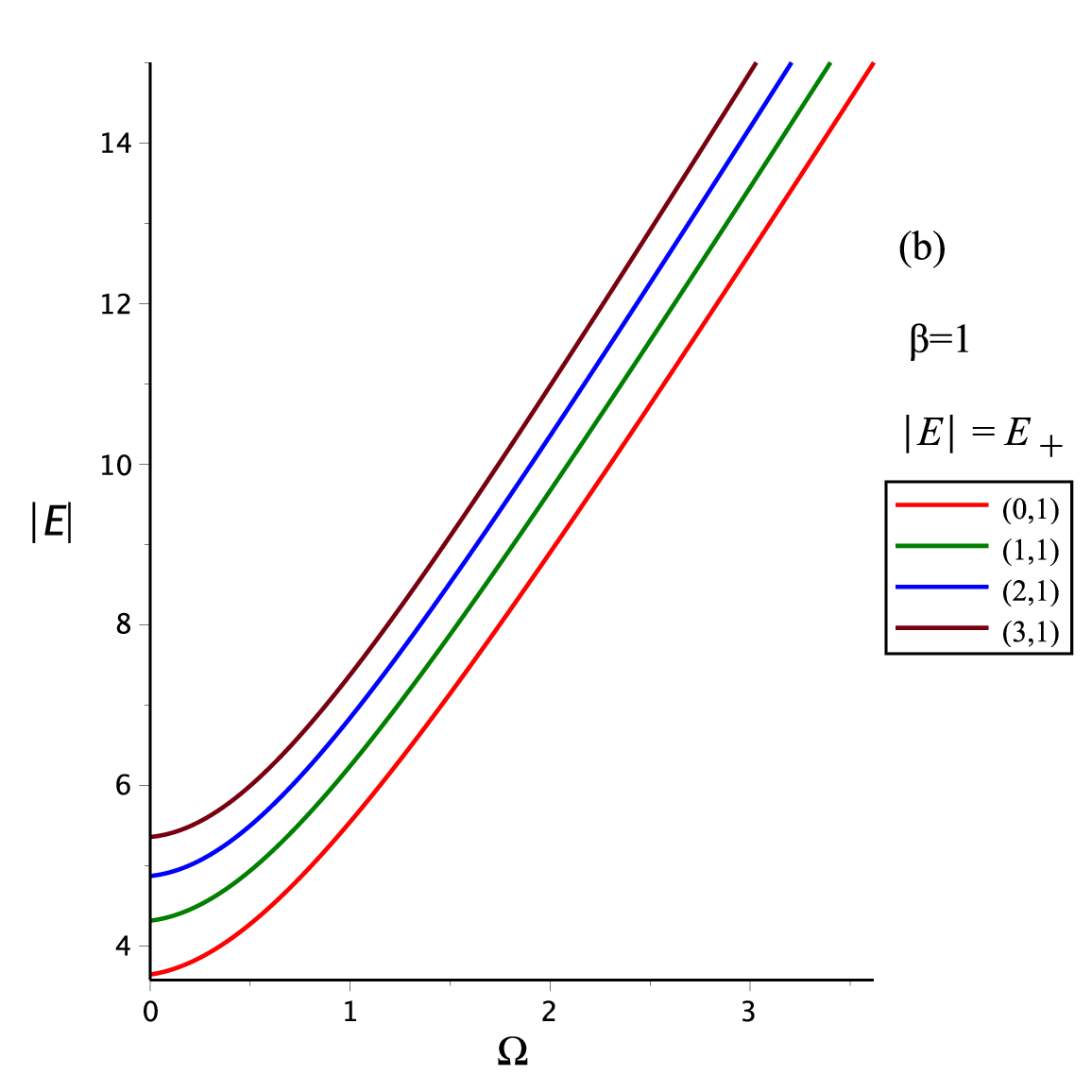}
\includegraphics[width=0.3\textwidth]{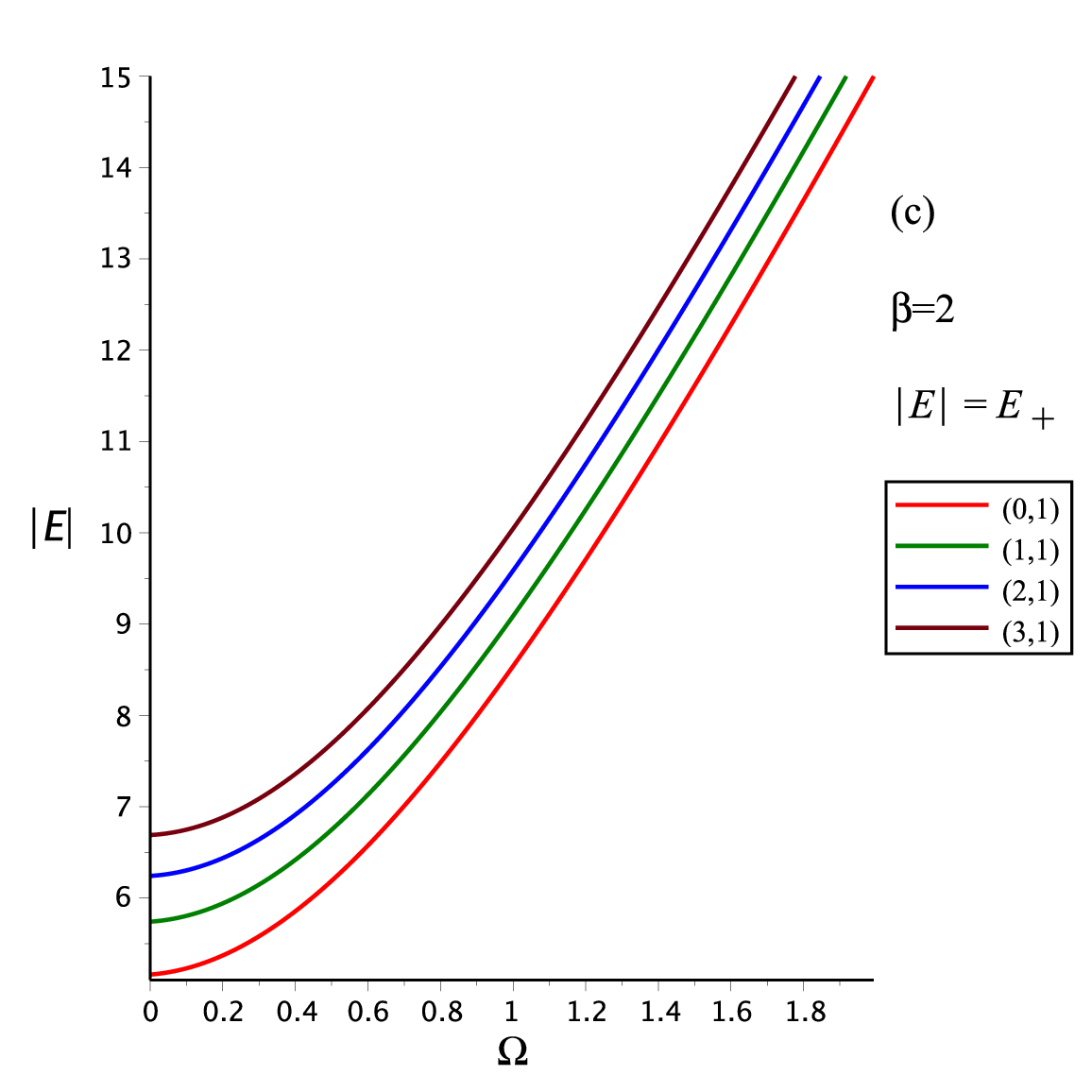}
\includegraphics[width=0.3\textwidth]{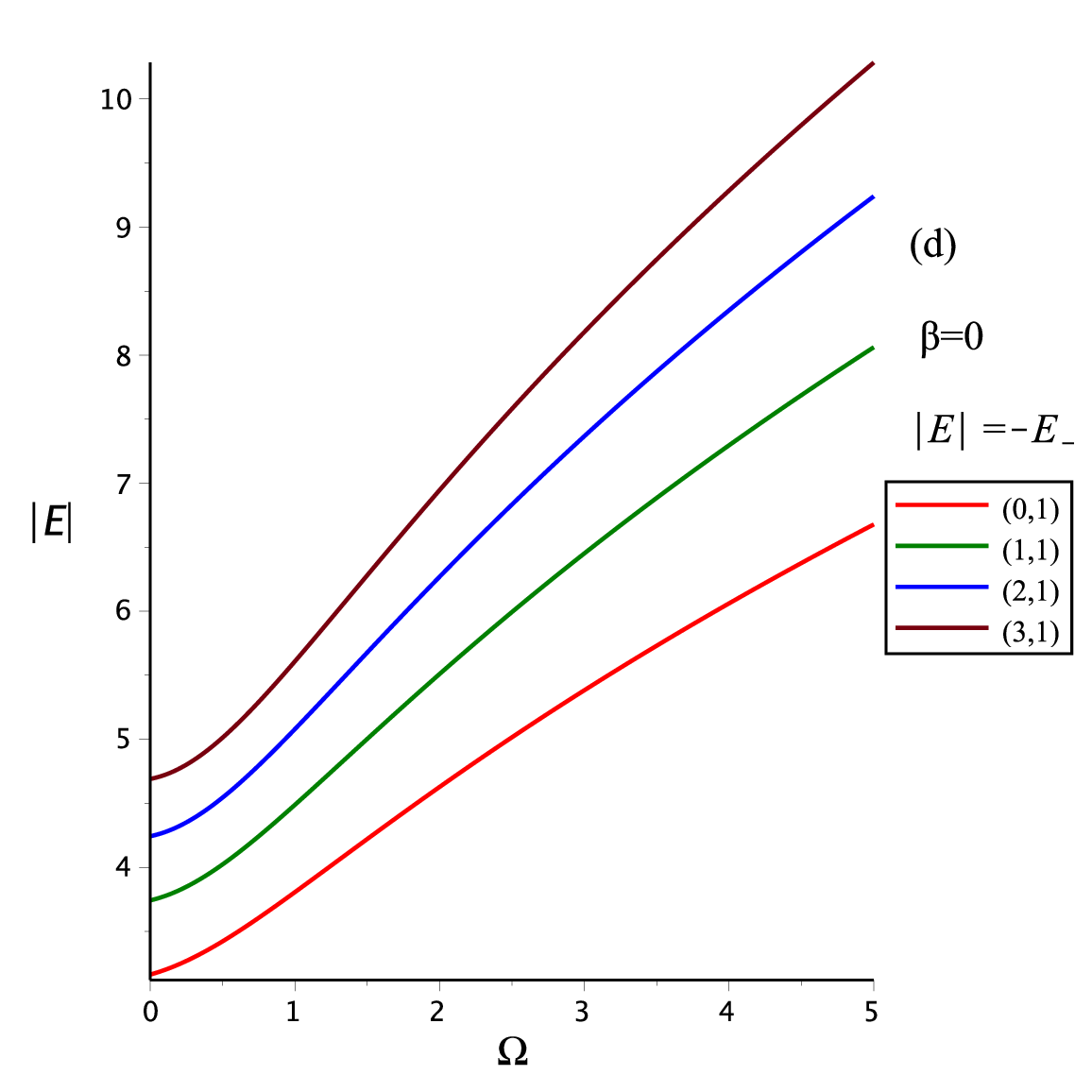}
\includegraphics[width=0.3\textwidth]{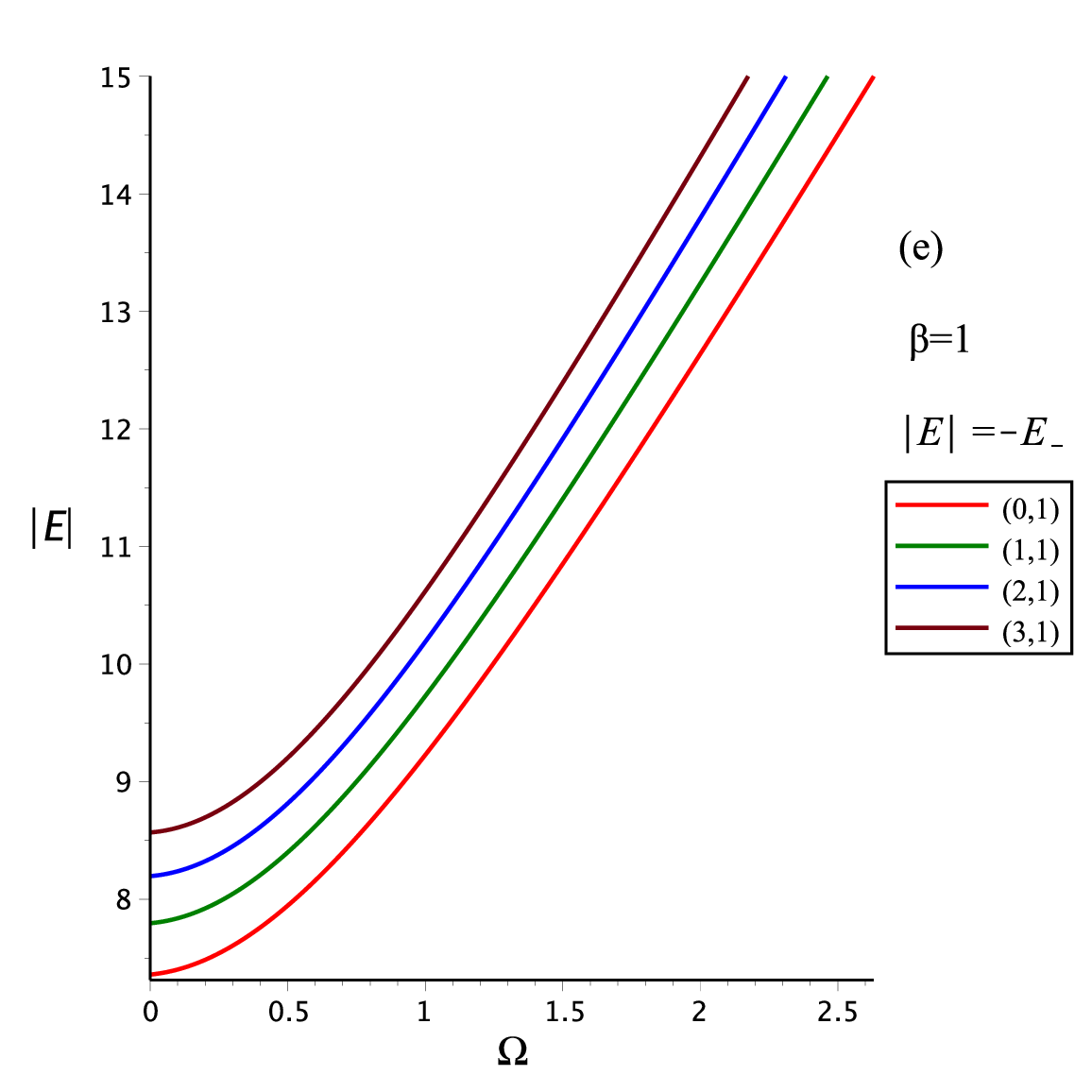}
\includegraphics[width=0.3\textwidth]{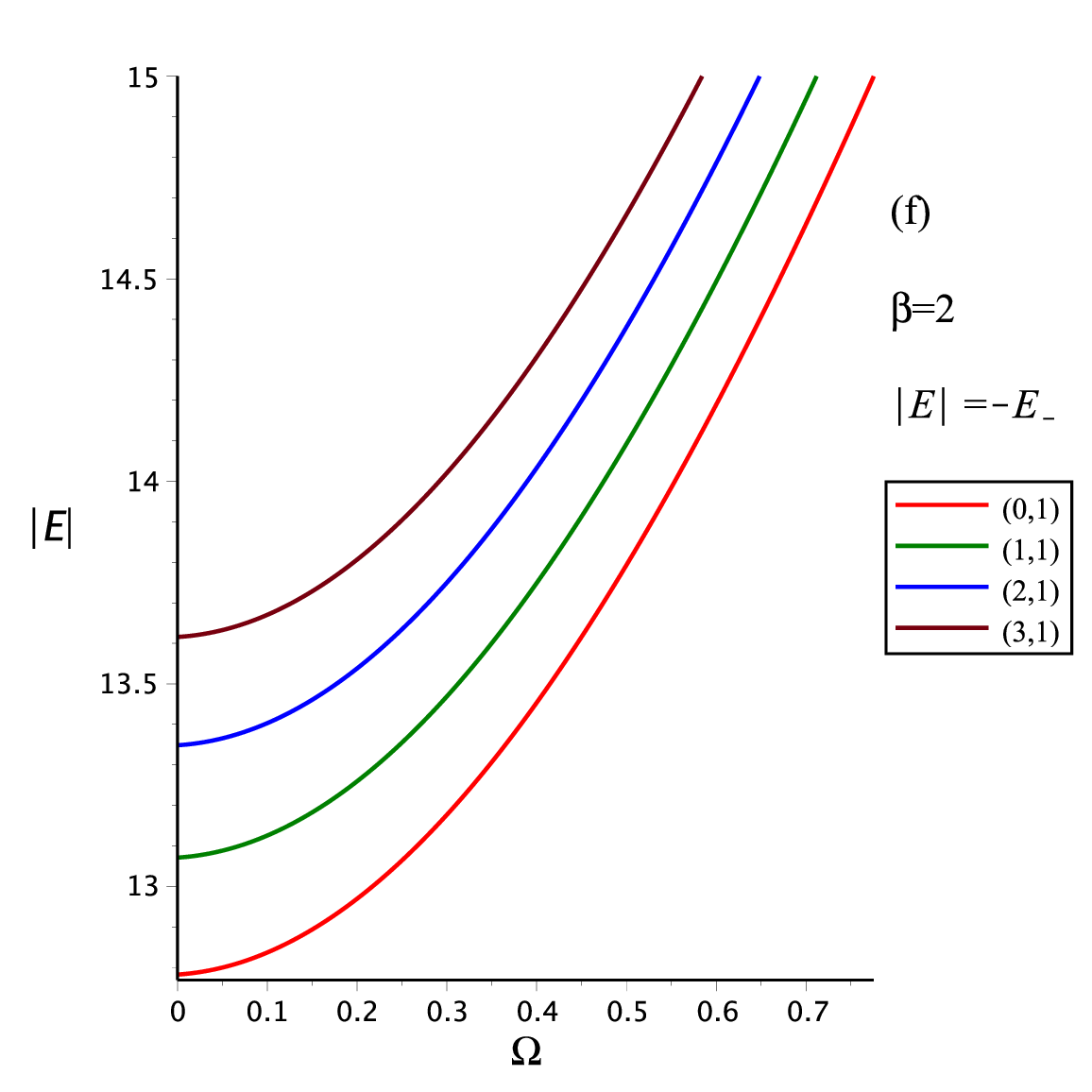}
\caption{\small 
{ The energy levels (\ref{2.19}) against the KG-oscillators'
frequency $\Omega $ (for $m=1\neq 0$ and $n_{r}=0,1,2,3$ at $\alpha =0.5$,
and $m_{\circ }=1=e=k=B$) so that the particles' energies are given in 4(a)
for $\beta =0$, 4(b) for $\beta =1$, and 4(c) for $\beta =2$. The
anti-particles' energies are given in 4(d) for $\beta =0$, 4(e) for $\beta =1
$, and 4(f) for $\beta =2$.}}
\label{fig4}
\end{figure}%

In Figures 3 and 4, we show in each figure the energy levels for particles and antiparticles, for the magnetic quantum number $|m|=1\neq 0$ and $%
n_{r}=0,1,2,3$, given in our result (\ref{2.19}) at $\alpha =0.5$, and $%
m_{\circ }=1=e=k$. Where, in Fig.3 we plot the energies against the magnetic field strength $B$ (at $\Omega =1$) so that the KG-oscillators particles' energies are given in 3(a) for $\beta =0$, 3(b) for $\beta =1$, and 3(c) for $\beta =2
$. In order to allow clear comparison between the effects of the spinning cosmic string on particles and antiparticles, we show the antiparticles energies in 3(d) for $\beta =0$, 3(e) for $\beta =1$, and 3(f) for $\beta =2$. Whereas, in Fig.4 we plot the energies against the oscillators frequency $%
\Omega $ (at $B=1$) so that the particles' energies are given in 4(a) for $%
\beta =0$, 4(b) for $\beta =1$, and 4(c) for $\beta =2$. Whereas, the antiparticles energies are given in 4(d) for $\beta =0$, 4(e) for $\beta =1$, and 4(f) for $\beta =2$. Again, the common trend in all figures indicates that clustering of the energy levels for KG-oscillators (both particles and antiparticles)
is feasible for $\beta >>1$. 

\subsection{The energy sets for the case where $E=E_{\pm }$ and $m=m_{\mp }$}

This case would specifically identify a set of particles, $%
E=E_{+}=+\left\vert E\right\vert $, associated with the magnetic number $%
m=m_{-}=-\left\vert m\right\vert $, and a set of anti-particles, $%
E=E_{-}=-\left\vert E\right\vert $, associated with $m=m_{+}=+\left\vert
m\right\vert $, each set at a time. Under such settings, our result in (\ref%
{2.8}) would imply%
\begin{equation}
E_{\pm }^{2}+\frac{eB}{\alpha }\beta E_{\pm }\left( 1+\frac{m_{\mp }}{\beta
E_{\pm }}\right) -\frac{2\tilde{\Omega}\beta }{\alpha }\left\vert
E\right\vert \left\vert 1+\frac{m_{\mp }}{\beta E_{\pm }}\right\vert -%
\mathcal{G}_{n_{r}}=0.  \label{2.22}
\end{equation}%
To solve for this equation, one may introduce $\sigma =\pm 1$ and rewrite $%
\left\vert 1+\frac{m_{\mp }}{\beta E_{\pm }}\right\vert =\sigma \left( 1+%
\frac{m_{\mp }}{\beta E_{\pm }}\right) >0$. That is, the signature of $%
\sigma $ is determined by the requirement that $\sigma \left( 1+\frac{m_{\mp }%
}{\beta E_{\pm }}\right) >0$. Consequently, equation (\ref{2.22}) would read%
\begin{equation}
E_{\pm }^{2}+E_{\pm }\left( \frac{\beta }{\alpha }\right) \left[ eB\mp
2\sigma \tilde{\Omega}\right] -\mathcal{G}_{n_{r},m_{\mp }}=0,  \label{2.23}
\end{equation}%
to imply%
\begin{equation}
E_{\pm }=\frac{-\beta \left[ eB\mp 2\sigma \tilde{\Omega}\right] \pm \sqrt{%
\beta ^{2}\left[ eB\mp 2\sigma \tilde{\Omega}\right] ^{2}+4\alpha ^{2}%
\mathcal{G}_{n_{r},m_{\mp }}}}{2\alpha },  \label{2.24}
\end{equation}%
where 
\begin{equation}
\mathcal{G}_{n_{r},m_{\mp }}=\mathcal{G}_{n_{r}}-\frac{m_{\mp }}{\alpha }%
\left[ eB\mp 2\sigma \tilde{\Omega}\right] .  \label{2.25}
\end{equation}

Obviously, however, this case seems to be unfortunate in the sense that the determination of the signature of $\sigma $ depends on the determination of $%
\left\vert E\right\vert $ and vice versa.  That is, for the current case of having  $E=E_{\pm }$ and $m=m_{\mp }$, respectively, the energies for the KG-oscillators (particles and antiparticles) are left unfortunate for being indeterminable. This should also hold true for the special case of Cunha et al. \cite{1.19}. 
 
\section{Concluding remarks}

Motivated by the very few  studies (one on the non-relativistic \cite{1.20} and the other on relativistic \cite{1.19} quantum particles in a spinning cosmic string in a magnetic field, to the best of our knowledge), we have, in this work, studied the effects of a spinning cosmic string in an external magnetic field on the spectroscopic structure of KG-oscillators.  We have started with the KG-equation and brought it to the one-dimensional Schr\"{o}dinger-oscillator form that admits a solution in the form of the confluent hypergeometric functions. The textbook truncation of  the confluent hypergeometric function into a polynomial of order $n_r\geq 0$ allowed us to obtain the energies in terms of a quadratic equation (\ref{2.8}), which is delicate in nature. That is, equation  (\ref{2.8})  describes KG- particles and antiparticles energies, $E=E_{\pm }=\pm |E|$, along with the magnetic quantum number $m=m_{\pm }=\pm |m|$. 

The $S$-states (i.e., $m=0$) of the KG-oscillators in a spinning cosmic string spacetime and an external magnetic field are discussed.  We have reported the particles energies, $%
E=E_{+}$, associated with $m=m_{+}$ and the antiparticles energies, $%
E=E_{-}$, associated with $m=m_{-}$. We have observed that the coexistence of a spinning cosmic string and a magnetic field affects the particles and antiparticles in different ways. That is, the antiparticles energies  increase more rapidly than those of the particles. Moreover,  the spinning of the cosmic string seems to cause clustering of the energy levels for particles and more rapid clustering for the antiparticles energy levels. This would suggest that there are no distinctions between energy levels as the spinning parameter $\beta$ increases, especially when $\beta>>1$.  We have observed, nevertheless, that the KG-oscillators energies $E=E_{\pm }$ that are, respectively, associated with $m=m_{\mp }$ are left unfortunate.

However, by setting $\Omega=0$, one would retrieve the case considered by Cunha et al. \cite{1.19}. Where we observe that while the Landau-like energies of the KG-particles ($E=E_{+}$), with $m=m_{+}$, have no explicit dependence on the spinning string parameter $\beta$ or the wedge parameter $\alpha$, the KG-antiparticles ($E=E_{-}$), with $m=m_{-}$, have such an explicit dependence (as documented in (\ref{2.21})). This should, interestingly, indicate that the coexistence of a spinning cosmic string and an external magnetic field cancels out the effect of the wedge parameter $\alpha$ for KG-particles  ($E=E_{+}$), with $m=m_{+}$, but not for the KG-antiparticles  ($E=E_{-}$), with $m=m_{-}$.  Yet, the result in (\ref{2.21}), enforces the coexistence of both a nonzero spinning parameter $\beta$ and a non-zero magnetic field $B$ in order to be able to observe the string spinning effect on the KG-antiparticles. However, for $\Omega\neq 0$ with the magnetic field switched off (i.e.,  $B=0$), one obtains Landau-like levels for KG-oscillators' in a spinning cosmic string as documented in (\ref{2.21.1}). In this case, we observe the KG-oscillators energies are symmetric about $E=0$. To the best of our knowledge, the current study has not been reported elsewhere. Yet, unavoidably, it invites a question of delicate nature, in the process, as to "what would be the effects of the \textit{other-way-spinning} of the cosmic string on such KG-oscillators? A question that would need detailed contemplation of its existential validity.

\section{Appendix: Power series solution for (\ref{2.5})}
In this section, we substitute
\begin{equation}
 U(r)=e^{-\frac{\tilde{\Omega}r^2}{2}} F(r),
 \label{A1}
\end{equation}%
in (\ref{2.5}) to obtain
\begin{equation}
r^2 F^{''}(r)-2\tilde{\Omega}r^3 F^{'}(r)+\left[(\mathcal{\tilde{E}}^{2}-\tilde{\Omega})r^2-(\tilde{m}^2-1/4)\right] F(r)=0. \label{A2}
\end{equation}%
The change of variables  $y=r^2$ would yield 
\begin{equation}
4y^2 F^{''}(y)+(2y-4\tilde{\Omega}y^2)F^{'}(y)+\left[(\mathcal{\tilde{E}}^{2}-\tilde{\Omega})y-(\tilde{m}^2-1/4)\right]F(y)=0. \label{A3}
\end{equation}
We may now use the power series expansion 
\begin{equation}
    F(y)=\sum_{j=0}^\infty A_j \, y^{j+\nu}, \label{A4}
\end{equation}%
in (\ref{A3}) to imply
\begin{gather}
\sum\limits_{j=0}^{\infty }\left\{ A_{j}\,\left[ \mathcal{\tilde{E}}^{2}-\tilde{\Omega}-4\tilde{\Omega}(j+\nu) \right] +A_{j+1}\left[ \left( j+\nu
+1\right) \left( j+\nu +\frac{1}{2}\right) -(\tilde{m}^2-1/4)\right]  \right\} y^{j+\nu +1}  \notag \\
 + A_{0}\left[ 4\nu^2 -2\nu -(\tilde{m}^2-1/4)\right]  y^{\nu }=0.\label{A5}
\end{gather}%
Since $A_0\neq 0$, we take $ 4\nu^2 -2\nu -(\tilde{m}^2-1/4)=0$ to imply that $\nu=\frac{1}{4}\pm \frac{|\tilde{m}|}{2}$. We take $\nu=\frac{1}{4}+ \frac{|\tilde{m}|}{2}$ to secure finiteness of $F(y)$ at $y=0=r^2$ for $\beta=0$. Consequently, (\ref{A5}) reduces to
\begin{equation}
 A_{j}\,\left[ \mathcal{\tilde{E}}^{2}-\tilde{\Omega}-4\tilde{\Omega}(j+\nu) \right] +A_{j+1}\left[ \left( j+\nu
+1\right) \left( j+\nu +\frac{1}{2}\right) -(\tilde{m}^2-1/4)\right] =0. \label{A6}
\end{equation}%
One may now truncate the power series to a polynomial of order $n_r\geq 0$ through the condition that $\forall j=n_r$ we take $A_{n_r+1}=0$ and $A_{n_r}\neq0$. Under such conditions, we get 
\begin{equation}
   A_{n_r}\,\left[ \mathcal{\tilde{E}}^{2}-\tilde{\Omega}-4\tilde{\Omega}(n_r+\nu) \right]=0 \Rightarrow \mathcal{\tilde{E}}^{2}-\tilde{\Omega}-4\tilde{\Omega}(n_r+\nu) =0. \label{A7}
\end{equation}%
Which would in turn give, with  $\nu=\frac{1}{4}+ \frac{|\tilde{m}|}{2}$,  
\begin{equation}
 \mathcal{\tilde{E}}^{2}=2\tilde{\Omega}\left(2n_r+|\tilde{m}|+1\right),   
\end{equation}%
which is in exact accord with our result in (\ref{2.7}) above.

\textbf{Data availability statement:} 
The authors declare that the data supporting the findings of this study are available in the article. 

\textbf{Declaration of interest:}
The authors declare that they have no known competing financial interests or personal relationships that could have appeared to influence the work reported in this paper.

\end{document}